# Irradiation-induced Ag nanocluster nucleation in silicate glasses: analogy with photography

***R. Espiau de Lamaestre***[1,2]***, H. Béa***[1,+] and ***H. Bernas***[1,*]***,***

[1]CSNSM-CNRS, Université Paris Sud XI, 91405 Orsay, France

[2] Fontainebleau Research Center, Corning SA, 77210 Avon, France

and

***J. Belloni*** and ***J.L. Marignier***

Laboratoire de Chimie Physique, UMR CNRS/UPS 8000, Bât. 349, Université Paris Sud, 91405 Orsay, France

***Abstract***

The synthesis of Ag nanoclusters in sodalime silicate glasses and silica was studied by optical absorption (OA) and electron spin resonance (ESR) experiments under both low (gamma-ray) and high (MeV ion) deposited energy density irradiation conditions. Both types of irradiation create electrons and holes whose density and thermal evolution – notably via their interaction with defects – are shown to determine the clustering and growth rates of Ag nanocrystals. We thus establish the influence of redox interactions of defects and silver (poly)ions. The mechanisms are similar to the latent image formation in photography: irradiation-induced photoelectrons are trapped within the glass matrix, notably on dissolved noble metal ions and defects, which are thus neutralized (reverse oxidation reactions are also shown to exist). Annealing promotes metal atom diffusion, which in turn leads to cluster nuclei formation. The cluster density depends not only on the irradiation fluence, but also – and primarily **-** on the density of deposited energy and the redox properties of the glass. Ion irradiation (i.e., large deposited energy density) is far more effective in cluster formation, despite its lower neutralization efficiency (from $Ag^+$ to $Ag^0$) as compared to gamma photon irradiation.

***PACS numbers: 61.46.Hk, 61.80.Jh, 68.55.Ac, 61.43.Fs***

[+] present address: Unité Mixte de Recherche CNRS-Thales and Université Paris-Sud XI, RD 128, 91767 Palaiseau, France

[*] corresponding author: bernas@csnsm.in2p3.fr

**Key words :** silicate glasses, silver nanoclusters, ionizing radiation

## 1. Introduction

The coloring of glasses by the controlled introduction of appropriate metallic nanoclusters is one of the older techniques known to humankind. Although their optical properties were related to the existence of nanoclusters [1], and analyzed long ago [2], [3], determining the mechanisms which lead to the nucleation and growth of these nanoclusters is an ongoing scientific and technological challenge for glass-making [4] as well as – more recently – for potential applications to photonics [5], [6]. Research on precipitation control has involved many techniques [7]. Among these, methods involving the use of ionizing radiation (photons, electrons, ions) have repeatedly drawn interest because control was envisaged, via the irradiation fluence, over the amount of precipitated metal, the density and average size of the metallic nanoclusters. For example, adequate control was indeed observed via photon irradiation in so-called photosensitive glasses [8], and rather precise control via ion irradiation was recently demonstrated [9]. However, in none of these cases was the microscopic mechanism of ionizing radiation-induced precipitation elucidated. The present work aims at improving our understanding of these processes.

As summarized in § 2.1, the fact that electrons produced by photons or ionizing radiation may control metal nanocrystal growth in various media is now well known. The best and oldest example is the photographic process [10] in which visible light impinging on AgBr crystals forms a so-called latent image: photon absorption creates electron-hole pairs, whose components migrate and may be trapped on different sites, including defect centers or silver ions in different charge states. The efficiency of the clustering process depends on the competition between electron scavenging by $Ag^+$ ions and the electron recombination with holes, hence on the redox interactions within the photographic emulsion. For example, the sensitivity of photographic plates was enhanced tenfold by adding an efficient hole scavenger [11a]. The main role of the developer is to transfer electrons to $Ag^+$ in the vicinity of a stable cluster [11b].

Radiolysis studies [11b, 12] have contributed significantly to an understanding of nucleation and growth processes in aqueous solutions containing dissolved charged or neutral Ag oligomers, and the experimental work and the presentation of this paper attempt a somewhat similar approach. By combining optical absorption (OA) and electron spin resonance (ESR), we first show how gamma-ray irradiation (i.e., with comparatively low deposited energy densities) affects nucleation and growth control of metal nanocrystals in glasses via the redox property modifications induced by ionizing radiation-induced electrons and holes, in strong analogy with the photographic process. We have also studied the effect on defect creation and subsequent metal solute precipitation of large deposited energy densities (due to ion irradiation), although it is more difficult to detail because sample thicknesses (hence analysis sensitivity) are limited by the irradiating particle range. By performing experiments with both silicate glasses and silica hosts, we show that these nonequilibrium deposited energy density (DED) effects combine with the equilibrium redox properties of the host to determine the silver clustering behavior.

In order to put this work into perspective, we first present a brief review of the literature on irradiation effects in metal cluster precipitation in glasses, highlighting the knowledge available as regards microscopic precipitation mechanisms.

## 2. Metal cluster formation by photons and ions: a brief summary

As will be seen later, the precipitation mechanism and its efficiency differ considerably depending on the local density of electrons and holes created by the irradiation (which in turn is determined by the deposited energy density) and on their subsequent diffusion and recombination processes. Hence our separation of results concerning photon versus ion irradiation.

### 2.1 Photosensitive glasses (photon irradiation)

Precipitation control of metallic nanoclusters in glasses using ultraviolet (UV) photons was initiated in the middle of the last century by Dalton [13] for copper-doped glasses, by Armistead [14] and Badger et al. [15] for silver-doped glasses and by Stookey [16] for gold-doped glasses. Such glasses are referred to as photosensitive glasses in the literature. In contrast to glasses used in the present work, they usually contain a chemical reducing agent ($As_2O_3$ or $Sb_2O_3$) and a photosensitizer ($CeO_2$) - the latter enhancing the effect of low energy photon (typically 4 eV) irradiation by easy dissociation and electron emission. The noble metal precipitation requires that the metal ions initially contained in the glass be reduced or neutralized into metal atoms. Stookey [16] first suggested an analogy to the photographic process as known at that time: photoelectrons induced by irradiation were supposed to be trapped in the glassy network (latent image); subsequent annealing allowed them to diffuse and reduce metal ions in the glass; the neutralized species could then move in turn, nucleate and grow. Despite this remarkable intuition, many questions remained unsolved as to the nature of so-called photoelectrons, the trapping mechanism in the glassy network and the nature of the traps. Also, what about photo-holes and possible metal ion neutralization during the irradiation itself? Later work by Maurer [8] clearly demonstrated proportionality between the final cluster density and the irradiation fluence, but did not elucidate the microscopic aspects of nucleation and growth notably as regards the neutralization stage. Kreibig [17] found evidence for neutralization in OA experiments, but only 1 % of the initial $Ag^+$ was neutralized by UV irradiation of a photosensitive glass, in agreement with prior ESR experiments on gamma-irradiated metaphosphate glasses [18], X-ray irradiated silicate glasses [19] or gamma-irradiated silicate glasses [20], thus partially contradicting the mechanism proposed by Stookey. On the basis of his observations, Kreibig proposed that: (i) nucleation first proceeds by aggregation of neutralized silver (thus accounting for the proportionality between the cluster density and the irradiation fluence); (ii) growth occurs by $Ag^+$ diffusion to the surface of such clusters, where $Ag^+$ could be neutralized by reducing agents (As, Sb) if these were present in the glass. This process assumes homogeneous nucleation. However, Kreibig noticed that some growth could also occur in the absence of reducing agents. Moreover, he did not deal with the role of irradiation defects. The present work provides a study of these features.

**2. 2. Ion irradiation**

The discovery of ion irradiation-induced precipitation of noble metals in glass was actually a side-effect of Maxwell-Garnett's theory of light absorption in heterogeneous media [2], a century ago [21]. Ion irradiation effects on metal ion-containing glasses have been reconsidered since the 1970's, with many attempts to control supersaturation, nucleation and growth. A detailed review by the most active group in the field [22] summarizes the corresponding research and we refer to it for complementary information. A common feature of much work in the area so far was its interpretation in terms of the specific complexities of ion beam interactions with glass. This was justified by the existence of the two, quite distinct, mechanisms of ion beam slowing-down – so-called electronic stopping (inelastic interactions with target electrons) versus so-called nuclear stopping (elastic collisions with target atoms) – leading to quite different energy deposition processes in the target. The consensus today is [23] that directly or indirectly, both ion stopping mechanisms lead inter alia to the production of electron-hole pairs which may interact with defects and impurities in insulators. The experimental results we obtained in this regard have been discussed elsewhere [24]: our purpose in the present paper is to parallel the discussion of photon-induced nanoclustering summarized above, with an emphasis on the elementary interactions.

There has been surprisingly little systematic work on the microscopic mechanisms responsible for ion irradiation triggering of metal precipitation. Very early on, Arnold and Vook [25] showed that ionic silver contained in metaphosphate glass is neutralized when ion-irradiated in the electronic stopping regime. This is similar to the neutralization effect by chemical reduction of photon irradiation discussed above. However, no link was established between the neutralized quantity and the final cluster density (or total precipitated amount) of Ag. The very extensive work of the Padua–Venice group [22], as well as of other authors [26], [27], [28] has largely centered on the high-fluence ion implantation or irradiation synthesis of single- or double-component (core-shell) metallic nanoclusters, i.e., concentrations well above 1%, for which clustering mostly occurs without further treatment. Although such concentrations are obviously of potential interest for many applications, notably in optical guiding and switching, they are generally not those that provide the clearest information on elementary processes. Complications are due, e.g., to high concentrations of moving metallic ions and defects during implantation that may lead to radiation-enhanced or - induced diffusion. Also, collisional recoiling of light (notably oxygen) atoms modify the glass composition and lead to strong chemical potential gradients in and around the implanted profile depth. The latter is likely the main consequence of the existence of the so-called nuclear (i.e., collisional) stopping power of ions. Examples of this are found, for example, in Ref. 22 where Secondary Ion Mass Spectrometry (SIMS) and/or Transmission Electron Microscopy (TEM) reveal nanocluster depth distributions that are well outside the implantation profiles; there are also many examples of lognormal nanocluster size distributions which testify [29] to the complex interference of several processes in nanocluster growth.

A recent attempt [30] to obtain more information by irradiating exchanged Ag/Na silicate glasses was blemished by the Ag concentration being so high that clustering had already started in the unirradiated sample, as shown by a photoluminescence signal ascribable to $Ag_3^{2+}$. Valentin *et al.* [9] demonstrated the control of Cu nanocrystal density in initially $Cu^+$(1-2%)-doped silicate glasses. After a low-fluence ion irradiation ($10^{12}$ – $10^{13}$ $Br^{7+}$.cm$^{-2}$ at 12 MeV) in the electronic energy deposition regime, no Cu nanoclusters were detected by high resolution TEM. Annealing the samples in the 550–700 K range allowed controlled growth of a nanocluster population whose density depended in a simple way on the energy deposition statistics, and the long-term limit of the size distribution closely followed the Lifshitz-Slyozov-Wagner (LSW) [31] characteristics of Ostwald ripening. This result indicates that the primary irradiation has two effects: (i) it initiates nucleation (at a sub-observational level as in a photographic latent image) with an efficiency depending on the deposited energy density, and (ii) it modifies the initial solute $Cu^+$ ion population in such a way that growth may occur by diffusion of, e.g., $Cu^0$ atoms. The first of these effects is established by the agreement of the size distribution with LSW, due to the fact that the initial population of aggregates constitutes the total growth mass. The second effect was deduced from the very existence of thermally activated growth, but was neither quantified nor understood in detail.

### 2. 3. Methodology of the present work

Here, we have approached the nucleation and growth problem in a more systematic way. First and foremost, we performed $\gamma$ irradiations in order to determine the effect of redox chemistry on the precipitation processes. ESR and OA studies of gamma-irradiated silver-doped silicate glasses demonstrate that the various silver redox states and their interaction with the host, particularly with the electrons and holes generated by the irradiation, play a major role throughout nucleation and growth. The nucleation and growth sequence that we observe in our experiments is basically similar to those observed during radiolysis of $Ag^{+}$-containing aqueous solutions [11], supporting our interpretation of the experimental results in terms of the photographic process. The knowledge acquired on photon-assisted clustering provides a basis for extending our interpretation of the results in terms of redox effects to ion irradiation-induced precipitation (here, electron transfer to metal ions at high deposited energy densities).

Two significant differences between photography and glass hosts are (a) the role played in the latter by irradiation-induced defects that trap electrons or holes which are released upon annealing. This feature was important in our work, as revealed by the use of techniques (OA and ESR) that are particularly sensitive to it, and (b) the fact that both the host redox properties and the beam DED affect nanoclustering. The latter point was studied by comparing nanocluster formation in Ag-containing silicate glasses and silica. For clarity,

**Table 1** summarizes the different sets of experiments and the information deduced from them.

## 3. Experimental

Silver-doped glasses were studied in this work because the surface plasmon resonance (SPR) absorption that signals Ag clustering is intense and spectrally isolated from the interband transitions (as opposed to the cases of Cu or Au) [32]. The existence of Ag nanocrystals containing more than about 10-20 atoms may then be deduced from the SPR absorption signal, whose position coincides [33] with the value predicted from Mie theory. Furthermore, we will see below that several ionized or neutral silver oligomers are also observable in ESR and optical absorption, providing vital information on the initial stages of clustering. Either high-purity silica, or the simplest sodalime silicate base glass – composition 74 $SiO_2$, 16 $Na_2O$, 10 CaO (mol %) – were used in our experiments. Most of our studies were performed on chemically doped samples by adding $AgNO_3$ to the glass melt. The reaction $AgNO_3 \rightarrow Ag_2O$ occurred, as detailed in Ref. 34. Since our aim was to compare directly the degree of $Ag^+$ reduction due to different ionizing radiation densities, no reducing agent was added to the glass composition. The initial $AgNO_3$ content was adjusted so as to avoid metal precipitation during glass synthesis or subsequent annealing, as controlled by OA (**Fig. 1**) - Ag was thus dissolved as $Ag^+$ ions in the pristine glass. As also shown in Fig. 1, high temperature annealing alone did not lead to $Ag^+$ reduction. The solubility limit corresponding to this procedure restricted the maximum Ag concentration to 150 appm (100 appm corresponds to $10^{-2}$ mol $l^{-1}$). The glass samples were melted in a silica crucible at 1800 K during 3 hours, poured onto a steel plate and annealed at 30 K above the glass temperature $T_g$ = 823 K in order to remove strain. The silver concentration was then measured by dissolution and OA spectrometry. One mm-thick samples were cut and polished on both sides. We studied the effect of the host composition (notably its redox properties) on DED-assisted precipitation by comparing nanoclustering in irradiated silica and silicate glass. The latter's composition was chosen so that its basicity was higher than that of pure silica. For these studies, Ag was introduced into the samples by ion implantation (using the CSNSM ion implanter IRMA) in order to increase the Ag dopant concentration, thus allowing OA monitoring of Ag clustering. In the latter experiments, special care was taken to check possible differences between the consequences of annealing alone versus the sequence (irradiation + annealing).

**Table 2** summarizes the features (energy, flux, dose or fluence and penetration depth) of the irradiations performed on our samples. We have also listed the deposited energy density or, equivalently, the linear energy transfer as used in the $\gamma$-irradiation literature. Gamma-ray irradiations were performed (using the LCP-Université Paris Sud facility) at room temperature or at 80 K, with a $^{60}Co$ source emitting 1.33 and 1.17 MeV photons interacting with the glass via the Compton effect at a flux of 3 $kGy.hr^{-1}$. Ion irradiations with 11.9 MeV $Br^{7+}$ or 1.6 MeV $He^{+}$ beams were performed at room temperature with the CSNSM ARAMIS accelerator [35]. Ion currents were kept below 200 nA for $Br^{7+}$ and samples were clamped to a metal holder in order to avoid significant heating. All sample (30 min) anneals were performed in a quartz tube oven under a dry $N_2$ flux, at temperatures ranging up to 1100 K.

A Dual Beam Cary 500 Spectrophotometer was used for optical absorption (OA) measurements. The latter can identify defects and several Ag oligomers, as well as Ag nanoclusters via their SPR absorption. Spectra were corrected by subtracting the unirradiated glass absorption contribution. In the following, the spectra obtained after gamma irradiation are presented in terms of the optical absorbance, since the entire sample thickness was affected by the irradiation. On the other hand, ion irradiation only induced modifications in a thin layer of our samples, so that the corresponding results can only be presented in terms of the measured optical absorption. X-band ESR spectra were acquired at 120 K with a Brucker EMX Spectrometer (ER 041 XG Microwave Bridge) at a working frequency of 9.42 GHz. The species detected by ESR in such samples are either glass defects, electrons trapped on threefold-coordinated silicon (E') [36], Non-Bridging Oxygen Hole Centers (NBOHC) [37], or silver-related species such as $Ag^0$, $Ag^{2+}$, or $Ag_2^+$ ($Ag^+$ is undetectable because of its paired electrons). The g-values of all these species are close to 2. Spectra were therefore acquired by scanning the magnetic field between 2750 and 4250 G, the amplitude and modulation frequency being respectively 3 G and 100 kHz. The power was limited to 1 mW in order to avoid signal saturation.

Modifications of the silver concentration depth profile were monitored by SIMS (the low silver concentration precluded the use of Rutherford backscattering). SIMS analyses were performed with a Cameca IMS 4F probe (at LPSC-CNRS, Bellevue) using a $Cs^+$ primary beam and signal detection via an electron multiplier. The detection threshold was about 0.1 appm for silver. The $^{30}Si$ signal was simultaneously recorded in order to determine the surface position and detector efficiency. Depth calibration was achieved by measuring the depth of the post-analysis crater with a Tencor Stylus profilometer, assuming a constant sputtering rate. These concentration profile studies were necessary for the following reason. Among the different valence states of silver (mainly $Ag^0$ atoms and $Ag^+$ ions) existing in oxide glasses, only those which are mobile may contribute to Ag clustering. The oxidized form $Ag^+$ is known [38] [39] to be very mobile. Doremus [40] found a very high $Ag^+$ diffusion coefficient ($\sim10^{-8} cm^2 s^{-1}$) at 773 K in sodalime glass; no direct measurement of the diffusion constant for $Ag^0$ in silicate glasses is available to our knowledge, probably because of its very low solubility. The corresponding numbers are usually evaluated very indirectly [41] by adjusting growth properties to classical growth models: values as low as $10^{-13}cm^2s^{-1}$ at 773 K were deduced in this way, but these evaluations may be in error since such models do not include the contribution of $Ag^+$ to Ag nanocluster formation. Uncertainties also beset an attempt [42] at a direct measurement of the $Ag^0$ diffusion coefficient in which a fraction of the silver precipitated during diffusion, so that the experimental result must differ from the intrinsic diffusion coefficient.

In order to evaluate Ag species mobility, we implanted Ag into our base sodalime silicate glass with the IRMA-CSNSM facility [43], at 480 keV to a fluence of $10^{15}$ ions.cm$^{-2}$, i.e., a maximum concentration of 500 appm. Samples were post-annealed for 1 h at temperatures between 486 K and 823 K, and total silver concentration profiles (including all oxidation states) were measured by SIMS. Annealing at temperatures above 623 K for 1 h resulted in a flattened concentration profile (**Fig. 2**), lowering the average Ag concentration to only 7 appm, and even to 0.3 appm after annealing at $T_g = 823$ K. This demonstrates fast diffusion of Ag species, both inwards and towards the surface. The double-peaked Ag profiles obtained at lower temperatures are indicative of Ag trapping, so that it is impossible to deduce a diffusion coefficient from the results. The OA spectra (not shown here) of these samples remained identical to that of the pristine glass, showing neither Ag nanocrystal plasmon absorption nor any identifiable Ag oligomer-related absorption peak. Thus most of the silver must be retained – and diffuse – in the oxidized $Ag^+$ state, which cannot be detected by OA and a negligible amount of Ag has been reduced. Electron-hole recombination induced by Ag implantation is more efficient than are reactions between electrons and Ag ions. In the following, we assume that both valences of silver (0 and +1) are mobile. Fig. 2 confirms Ag mobility at low annealing temperatures, but there is no evidence for a relation to classical diffusion. The splitting of the initially quasi-gaussian profile into two parts, one shifted towards the surface and the other in depth, may be due to implantation-related effects (defect creation and strain; changes in the oxygen content inside the implantation profile by collisional recoiling) that modify the $Ag^0 \leftrightarrow Ag^+$ equilibrium and the diffusion properties of both species and other oligomers.

## 4. Results on nanocluster nucleation

The deposited energy density (DED) along the ionizing particle track, also designated in the radiolysis literature as the average value of the Linear Energy Transfer LET (**Table 2**), affects the electron and hole density as well as the radiation–enhanced (or –induced) Ag diffusion. Cluster nucleation therefore involves (i) the metal atom/ion's charge state; (ii) the stability of the corresponding charge states; (iii) the diffusivity of different Ag species in the glass, and (iv) the stability of small Ag aggregates versus charge capture. The redox potential of all these entities and that of the glass matrix affects the clustering efficiency. This conclusion emerges from the results presented in this Section.

A specific trait of glasses is that the clustering sequence is mediated by a stage in which irradiation-induced defects act as charge reservoirs. The nature and stability of these defects therefore plays an important role in the analysis, and § 4.1 describes experimental results on this point. In § 4. 2, we describe our ESR and OA studies of Ag clustering induced by $\gamma$ photon irradiations – hence at a low deposited energy density ($^{60}Co$ gamma rays generate Compton electrons of about 600 keV, whose mean linear energy loss is about 0.5 $eV.nm^{-1}$). The Ag content of the doped glass was in the range 17 - 130 appm. The sensitivity of both ESR and OA to signals from defects and from Ag species in doped samples is high, the whole volume of the sample being modified by the irradiation, so that rather complete information on the initial stages of nucleation and growth is obtained. We also performed heavy ion (1.6 MeV He and 12 MeV Br) irradiations of the same base glasses containing 65 appm or 130 appm Ag: these irradiations involve very large DED (resp. ~ 0.3 $keV.nm^{-1}$ and ~ 4 $keV.nm^{-1}$) (**Table 2**). They have the experimental drawback of modifying the sample only at a near-surface depth (typically below 4 μm, the average ion penetration depth), hence at levels close to the ESR and OA detection sensitivity limits. Notwithstanding this limitation, the comparison of irradiations for which the DED values differ by 4 orders of magnitude demonstrates the effect of varying concentrations of electron-hole pairs in the glass.

### 4. 1. Charge reservoirs: irradiation-induced defect centers in undoped glasses

A guideline in this area is the considerable work on initial charge formation and evolution (the solvated electron) in aqueous solutions [12] [45]. The obvious difference here [23] [46] is the formation of metastable defects which "store" the charges in different ways. The influence of DED on the creation and short-term evolution of different electron- and hole-capturing glass defects was studied first by comparing the ESR spectra of γ- versus ion-irradiated glasses containing little or no Ag. Fig. 3 shows that ESR is quite sensitive to these defects, in spite of the ion-range limitation. We observed the trapping [44] of the photoexcited carriers in the glassy network via both ESR and OA. The so-called "non bridging oxygen hole center" (NBOHC) was previously identified [47], [48] as the source of the intense ESR signals at $g = 2.0097$ and $g = 2.0036$. Our spectra indicate that γ-irradiation led to a very large NBOHC population (Fig. 3a). We also found a small signal at $g = 1.964$, ascribed to a trapped electron [47], but we observed no ESR signal from E' or peroxide radical (POR) centers [46].

In the ESR spectra corresponding to ion irradiation (Fig. 3b), the NBOHC contribution remained approximately unchanged (i.e., it varied as the ratio of the irradiated sample thicknesses), but here the E' centers' contribution to the spectra was very large, demonstrating the impact of a large DED in their formation. The recombination and subsequent dissolution of these defect centers upon annealing at the known [46] E' defect temperature stage is clearly shown in **Fig. 4**. Potential signals due to electrons or electron traps were not intense enough to be detected by ESR in these samples.

Complementary information was obtained from OA, whose sensitivity to the different absorbing centers differs from that of ESR. The OA spectra of the $\gamma$-irradiated undoped silicate glass display overlapping absorption bands. Its spectrum was fitted here by a combination of gaussian curves whose parameters are given in **Table 3**: the resulting absorption bands are close to those previously found [47] in a similar glass. We detected the well-known NBOHC (1.98 and 2.79 eV bands) and now also observed the peroxy radical POR (5.35 eV band) hole-trapping centers (not seen in ESR spectra), as well as a known [47] electron-trapping center (4.1 eV band). The electron-trapping E' center, which absorbs around 7 eV (VUV), was not observable with our OA setup. **[see Note Added]**

### 4. 2. Silver species evolution after $\gamma$ irradiation of $Ag^+$ doped glasses

Silver-doped silicate glass samples (65 appm) were $\gamma$-irradiated to 15 kGy at 80 K and then annealed for 30 min at increasing temperatures. The corresponding ESR spectra (Fig. 5) provide considerable information on the initial stages of Ag species' formation and evolution. Up to 363 K, they displayed signals typical of $Ag^0$, [19], [20], [49] with the hyperfine splittings due to the electron-nuclear spin (I=1/2) interaction for both silver isotopes (abundances: 51.8 % for $^{107}Ag$, 48.2 % for $^{109}Ag$). Thus, a fraction of the photoelectron population has reduced silver ions directly by the reactions:

$$\text{Silicate glass + irradiation} \rightarrow e^-, \text{ NBOHC, POR defects} \quad (0)$$

$$Ag^+ + e^- \rightarrow Ag^0 \quad (1)$$

(the spectrum termed "unannealed", taken at 120 K, only showed inhomogeneous broadening – the isotopic silver doublet could not be observed since very slow structural relaxation prevented the silver environment from reaching its neutral silver configuration). The intensity of signals due to solvated electrons or other electron traps was too low for ESR detection.

Evidence for $Ag^0$ was also found by comparing the OA spectrum of the undoped glass to that of a $\gamma$-irradiated (40 kGy) 17 appm silver-containing glass. The latter (Fig. 6) displayed an additional absorption band centered at 3.6 eV ($\lambda$ = 345 nm), whose features were determined (Table 3) by a multigaussian adjustment as above, while keeping the peak position and width of the previously determined undoped glass bands fixed. The new band's position agrees with that found for free $Ag^0$ [50] as well as for $Ag^0$ in water (360 nm) [12]. The optical absorption of a species depends on the polarity of the medium in which it is embedded, and the nature of ligands that interact with it: the fact [51] that in oxide glasses $Ag^0$ is surrounded by oxygen, just as in water, likely accounts for this result. The concomitant identification of $Ag^0$ in ESR and OA confirms [17] that photon irradiation leads to the neutralization of ionic silver.

The observation of neutral silver means that $Ag^+$ traps electrons, but its role in the photocarrier population evolution is not a simple one, since all defect absorption intensities were two to five times lower in the doped glass than in its undoped counterpart (Table 3). The presence of silver must in fact also catalyze electron and hole recombination. Actually, $Ag^0$ can trap holes via the reaction:

$$Ag^0 + h^+ \rightarrow Ag^+ \qquad (2),$$

A reaction with a detrapped hole may also occur, such as:

$$Ag^+ + h^+ \rightarrow Ag^{2+} \qquad (3)$$

followed by a reaction of oxidized $Ag^{2+}$ with electrons :

$$Ag^{2+} + e^- \rightarrow Ag^+ \qquad (4).$$

Reactions (1) and (4) account for the electron trap population decrease when $Ag^+$ is added to the glass. The recombination center role of multivalent elements such as Ag was previously documented [52], but not insofar as its role in post-irradiation clustering is concerned. Such Ag-assisted recombinations may presumably take place along the Compton electron track (before electron- and hole-trapping), as well as by short-term diffusion of electrons and holes detrapped from the abovementioned centers. Our observations indicate that silver species interact with trapped charges at the earliest irradiation stage, as well as after annealing. To our knowledge, this early redox interaction had not been noted in previous work.

Other silver species were identified, and we monitored their concentration versus the irradiation defect concentration variations. As shown in Fig. 5, annealing the sample at 413 K led to the disappearance of the $Ag^0$ signal and the growth of a new doublet with a lower g-value. The latter is well-established [53], [54], [55] as characterizing $Ag_2^+$ ($g = 1.994$, $A = 780$ Mhz = 279 G). Thus, neutral silver has reacted with $Ag^+$:

$$Ag^0 + Ag^+ \rightarrow Ag_2^+ \qquad (5)$$

This also appeared in OA spectra taken on similar samples (**Fig. 7**) in which the absorption peak is blue-shifted from 345 nm (the characteristic wavelength of $Ag^0$) to 310 nm. We ascribe this 310 nm peak to $Ag_2^+$; its position is identical to that found for $Ag_2^+$ in radiolyzed aqueous solutions containing $Ag^+$. This identification is supported by the correlated reduction in the intensity of this OA peak and of the $Ag_2^+$ ESR signal in Fig. 5. Annealing at 413 K also led to an ESR signal with a uniaxially anisotropic g-factor ($g_{//} = 2.28$; $g_{\perp} = 2.04$). This feature is ascribed to $Ag^{2+}$ [56], [57] and provides evidence for the reactions:

$$Ag^+ + NBOHC \rightarrow Ag^{2+} + NBO \qquad (6)$$

and

$$Ag^+ + POR \rightarrow Ag^{2+} + \equiv Si - O - O^- \qquad (7).$$

The integrated ESR signals are proportional to the concentration of the various defects and Ag species observed in ESR. **Fig. 8** compares their temperature dependence for the 130 appm Ag glass, γ-irradiated to 15 kGy at 80 K. The corresponding intensities were deduced by double integration of the ESR signals. The intensity due to all observed defects, as well as to $Ag^0$, decreased as the temperature rose from 120 K to 290 K. Electrons and holes recombine by the reactions:

$$NBOHC + e^- \rightarrow NBO \qquad (8)$$

$$Ag^0 + NBOHC \rightarrow Ag^+ + NBO \qquad (9).$$

As mentioned previously, our experimental conditions did not allow detection of POR centers. Mechanisms (8) and (9) led to a decrease of the $Ag^0$ population as the temperature rose to 300 K, followed by a relative increase presumably due to a release of trapped electrons according to Equation (1). From 360 K to 420 K, the decrease in NBOHC and $Ag^0$ populations were correlated to an increase in that of $Ag^{2+}$ and $Ag_2^+$, respectively (Eqns. 5 and 6). The role of the NBOHC oxidizing center is critical in this evolution. The ESR signals diminished at temperatures above 425 K and disappeared above 580 K, but the OA spectra taken in the temperature range between 530 K and 630 K (Fig. 7) displayed a shift from 310 nm ($Ag_2^+$) to 280 nm. Paramagnetic defect centers (notably the NBOHC) are essentially annealed out at such temperatures according to the ESR results, so that the latter absorption peak is very likely due to an interaction between Ag species. A comparison with results obtained for Ag-containing aqueous solutions [12] is instructive again: in that case, an absorption peak was also found at 280 nm, due to $Ag_3^+$, $Ag_3^{2+}$, or to $Ag_4^{2+}$ formed by the diffusion and dimerization of $Ag_2^+$ ions. The latter process is unlikely in glass because $Ag_2^+$ mobility is low compared to that of $Ag^0$ and $Ag^+$, but Ag° or $Ag^+$ may diffuse to and react with the $Ag_2^+$ oligomer by the reactions:

$$Ag_2^+ + Ag^0 \rightarrow Ag_3^+ \qquad (10)$$

$$Ag_2^+ + Ag^+ \rightarrow Ag_3^{2+} \qquad (11),$$

so that we ascribe the weak 280 nm peak, found after γ-irradiation (to 21 kGy) and post-annealing at temperatures up to 643 K, to either $Ag_3^+$ or $Ag_3^{2+}$, or both. To our knowledge, these experiments are the first to detect the initial steps of metal aggregation in a silicate glass. They show that standard phase diagram considerations alone cannot account for nanocluster formation in glasses: the aggregation process also depends on the outcome of a multifactor competition between silver reduction and oxidation.

Silver reduction being a prerequisite for nanocluster formation, information on the $\gamma$ irradiation-induced neutralization efficiency was obtained in the following way. We first deduced the $Ag^0$- and defect-creation rate from the irradiation dose-dependence of the OA spectra. As noted previously [58], the evolution of the $Ag^0$ and defect populations is well described (**Fig. 9**) by the simplest rate equation solution:

$$y = 1 - \exp(-\rho D) \qquad (12)$$

where y is the defect fraction, D the $\gamma$ irradiation dose and $1/\rho$ a defect creation dose corresponding to the transition from the independent to overlapping track regime. Adjusting the experimental defect fraction in Fig. 9 gives $1/\rho \approx 7$ kGy. Since 1 Gy corresponds to $5\times10^{10}$ $\gamma$ interactions per $cm^{-3}$ of glass, we find that the effective defect creation volume per $\gamma$ interaction is approximately $10^{-15}$ $cm^3$. Considering the approximations made, this is acceptably close to the volume ($10^{-14}$ $cm^3$) of the Compton electron track as calculated by Cooper [59], so that the efficiency of silver reduction within the track may be roughly estimated from Fig. 9. We assume, in view of their similar environment and spectral properties, that the molar extinction coefficient per neutral silver atom in silicate glasses is identical to that in water, i.e., 17000 l $mol^{-1}cm^{-1}$ [12]. In the 118 appm silver-doped glass, a $\gamma$ irradiation dose of 21 kGy led to saturation of the Ag neutralization at an intensity of about 15 $cm^{-1}$, i.e. a concentration of $5.5 \times 10^{17}$ $Ag^0.cm^{-3}$. The initial concentration of $Ag^+$ being $7 \times 10^{18}$ $cm^{-3}$, we might conclude that some 10 % of the initial silver ion population was reduced by irradiation. In fact, this is actually a gross underestimate of the neutralization efficiency since the sample was irradiated and maintained at room temperature, and the optical spectrum was recorded five days after irradiation. Evidence that a significant fraction of $Ag^0$ had by then been reoxidized by recombination with trapped holes is provided by Fig. 5, which shows that a 30 min anneal at 295 K leads to more than a fivefold decrease of the $Ag^0$ signal relative to its intensity in the as-irradiated (at 80 K) sample.

We argue that silver neutralization is in fact largely achieved in the initial Compton electron track, but that reverse oxidation reactions such as (2) and (9) occur in time, affecting the ratio of the $Ag^0$ specie population to that of its charged counterpart. The yield G (ratio of the newly formed $Ag^O$ species concentration per unit of deposited energy, in SI units of molecules per Joule) may be estimated from the initial slope of the curve in Fig. 9. We measured about $2.2 \times 10^{17}$ $Ag^0$ $cm^{-3}$ ($3.5 \times 10^{-4}$ mol $l^{-1}$) at 5 kGy (after a 5-day anneal at room temperature), leading to $G = 0.7 \times 10^{-7}$ mol $J^{-1}$. This may be compared to the yield found [11] in aqueous solutions, i.e., $G = 2 \times 10^{-7}$ mol $J^{-1}$. Extrapolating our value to 80 K (no annealing, no reverse oxidation) would produce a yield of about $3.5 \times 10^{-7}$ mol $J^{-1}$, similar to that ($4 \times 10^{-7}$ mol $J^{-1}$) of the hydrated electron at very short times before electron-hole recombination may occur. It is noteworthy that effects which only appear at very short time scales in water become easily visible in glasses due to slow Ag species diffusion.

## 5. Results on clustering efficiency and growth

In this Section, we show that both the growth mechanism and the ultimate clustering efficiency are also strongly affected by the interaction of the neutral and ionized Ag species' interactions with the charge populations released from the electron- and hole-containing defects. The formation of nanoclusters in glass requires that the medium be sufficiently reducing to favor growth over dissolution, i.e., reactions (1) and (4) as well as reactions of $Ag^+$ with trapped electrons, over reactions (6) (7) (9) in the initial stages.

### 5.1. Growth after γ-irradiation

After annealing above 640 K, the OA spectra of the gamma-irradiated sample (**Fig. 10**, extending the results of Fig. 7 to higher annealing temperatures) display the Ag peak at 400 nm, i.e. the collective excitation (plasmon) wavelength as derived from the standard Mie calculation. The peak was shown, in the case of water, [60] to be fully developed when the clusters contained at least some 13 atoms on average. We estimated the total amount of precipitated Ag in our samples from the OA peak intensity, as in § 4.2, to be approximately $4\times10^{16}$ Ag $cm^{-3}$, i.e., about 1% of the total Ag concentration in the glass. This means that only some 10% of the neutralized Ag° finally coalesced into clusters. The actual proportion may have been somewhat higher, since we assumed that the OA extinction coefficient was 17000 l $mol^{-1}cm^{-1}$, whereas it is in fact known to vary from nearly zero for $Ag_4$ to 15000 for $Ag_{13}$ and a fraction of the clusters were probably smaller than $Ag_{13}$. But it is clear that a very large fraction of the initial $Ag^0$ population was reoxidized during the anneal, presumably by interacting with the hole population released from traps as the temperature was raised (e.g., Eqn. 9). Recall that annealing all the defects at temperatures above $T_g$ (see the OA spectra of Fig. 4) led to complete dissolution of all clusters. In our base glass composition – devoid of added reducing agents – the initial amount of reducing electron defects equals the initial amount of hole defects. At sufficiently high temperatures, detrapped holes may recombine not only with electrons but also with silver oligomers, thus oxidizing the latter.

The metastability of different Ag species was also shown in the following way. Ag (118 appm)-doped glass samples were room-temperature gamma-irradiated to a 21 kGy dose and annealed for 30 min, either at 373 K (most of the silver is then $Ag^0$) or at 413 K (most of the silver is then $Ag_2^+$ and $Ag_3^+/Ag_3^{2+}$). We compared (**Fig. 11**) their optical absorption immediately after the anneal and after room-temperature conservation for 11 months. In the former case, the $Ag^0$ contribution practically disappeared; in the latter, the $Ag_2^+$ concentration decreased by 20 % while that of $Ag_3^+$ or $Ag_3^{2+}$ remained practically identical. We conclude that $Ag^0$ is more easily oxidized than $Ag_2^+$, the reverse being true of $Ag_3^+$ and $Ag_3^{2+}$. This suggests that the redox potential increases with the nuclearity, as previously demonstrated [11] in water, and leads us to the following stability sequence for the various species:

$$Ag^0 < Ag_2^+ < Ag_3^+/Ag_3^{2+} < \ldots < Ag_n \quad (13).$$

It indicates the existence of a "critical" aggregate size (~3 Ag entities), below which clusters are unstable versus reoxidation (hence redissolution) due to holes released by glass defects, and above which stable nanocrystals may grow by $Ag^+$ diffusion to the cluster and *in situ* reduction by an electron.

### 5. 2. Growth after ion irradiation

Although not as detailed, information on the clustering mechanism under 1.6 MeV He- or 12 MeV Br- ion irradiation is quite significant. In contrast to previous [22], [61] ion beam-assisted metal colloid formation experiments, our purpose here was to distinguish initial effects specifically due to irradiation from those due to annealing-induced mobility. A prior study of the heat input by the ion beam was carried out [34] [24] in order to identify the former unequivocally. By keeping the dissipated power below 0.3 $W.cm^{-2}$, sample heating (hence Ag diffusion) remained negligible, and all observed effects were due to the large DED. The silicate glass's composition was identical to the one used in our $\gamma$-irradiation studies above, and it was chemically doped with silver to 130 appm. **Fig. 12** shows the OA spectra for such a glass after Br ion irradiation to $10^{14}$ ions.cm$^{-2}$ and post-annealing up to 693 K. The only significant feature in the spectra was the appearance of the characteristic $Ag_n$ plasmon frequency at 400 nm at temperatures as low as 570 K, with a maximum intensity after annealing to 643 K, and progressive dissolution as the annealing temperature approaches $T_g$. This result concurs with those obtained [9] on 12 MeV Br irradiation-induced clustering of Cu and Ni in similar glasses via this "ion beam photography" process. However, the results presented here for Ag clustering also show significant differences with those obtained in that work. One difference involves the use of experimental techniques with different size threshold detection limits (OA versus TEM), while others relate to the differing metal species' redox potentials and mobilities. A remarkable example of the latter is the ion irradiation fluence-dependence of the plasmon peak intensity after a 643 K anneal, shown in **Fig. 13**. These and other ion irradiation results are discussed elsewhere [24] in more detail. For our present purpose, we note that the curve saturation value corresponds to $4 \times 10^{15}$ atom cm$^{-2}$ (or $10^{18}$ atom cm$^{-3}$ in the 4 μm irradiation depth). Thus, about 15% of the entire silver content in the irradiated portion of the glass formed stable clusters after the 12 MeV Br irradiation, as opposed to the 1% value after γ-irradiation. While both reduction and oxidation occurred in the low-DED conditions of the latter, the high DED drastically modifies the nucleation and growth conditions. Why ? The 15-fold increase in the clustered Ag fraction does not signal a larger neutralization efficiency, since the ion fluences are some 4 orders of magnitude higher than the γ doses shown above. The radiolytic yield estimated from Fig. 13 was only $G = 0.4 \times 10^{-8}$ mol J$^{-1}$, i.e., 20 times lower than that of a γ-irradiation. A previous comparison [62] of the neutral silver yields after irradiation by a $C^{6+}$ ion beam or by γ-photons in an aqueous medium gave similar results. The reduced neutralization yields observed for high DED radiation versus those due to γ-photons in both media are presumably due to the efficient electron-hole recombination (reaction 8) in the plasma-like track environment, a recombination that occurs faster than silver reduction (reactions 1 and 4). Such reactions could also be the source of the observed enhancement in stable Ag clustering (see Section 6).

### 5. 3. Influence of host composition

The results presented so far show that Ag clustering is affected by the irradiation-induced local modification of the sample redox potential. The outcome of a given ion irradiation DED as regards Ag clustering should in turn depend on the initial redox state of the host. This is indeed the case, as shown by the following experiments. A concentration ~ 0.1 at% Ag was ion-implanted ($10^{15}$ Ag $cm^{-2}$, 480 keV), rather than dissolved, at a projected range $R_P \approx 160$ nm in either pure silica or silicate glass samples, i.e., in different redox environments. OA measurements were performed in order to detect $Ag^0$ and other oligomers, as well as silver clusters. In the case of the Ag-implanted pure silica samples (**Fig. 14, upper**), annealing at increasing temperatures led successively to Ag reduction, oligomer formation (at least $Ag_2$ and $Ag_3$ clusters) and nanocluster formation (detected by the SPR). By contrast, the OA spectra (not shown here) corresponding to the Ag-implanted silicate glass samples showed no evidence of Ag reduction or precipitation even after annealing to temperatures in the vicinity of the glass transition temperature (863 K) where species mobility is very high (see **Fig. 2**). We ascribe this absence to the fact that silicate glass is considerably less reducing than silica. The chemical nature (redox degree) of the glass host thus has a major impact on its sensitivity to irradiation. This was also apparent in a comparison of identical 12 MeV Br irradiations (fluence $2 \times 10^{14}$ Br $cm^{-2}$) performed on unannealed Ag-implanted silica versus silicate glass. The $Ag^+$-implanted region was confined to the high electronic DED regime part (4-5 keV/nm) of the Br ion tracks in both cases. The spectra obtained after high DED irradiation of Ag-implanted pure silica and annealing (**Fig. 14, lower**) were very similar to those obtained (Fig. 14, upper) on its unirradiated counterpart, i.e.: the high DED irradiation only had a weakly accelerating effect on nanoclustering, the silica host being sufficiently reducing to insure precipitation of the entire Ag content above 870 K. On the other hand, whereas no clustering had been found upon annealing the Ag-implanted silicate glass in the absence of the high DED irradiation, the latter had a major effect on the system evolution (**Fig. 15**). It reduced silver even at room temperature, and led to nanocluster formation (detected by the SPR at 400 nm) above ~ 500 K in the silicate glass. The interaction between the reducing power of ion-irradiation DED and the redox degree of the matrix was clearly substantiated by the occurrence of a maximum in the SPR intensity at an annealing temperature around 620 K. Below this temperature, the reducing action of irradiation DED counterbalanced the oxidizing action of the matrix, whereas above 620 K the thermodynamical state of silver in the silicate glass favored Ag oxidation, leading to cluster dissolution. In summary, efforts to control nanocluster formation require that the redox properties of both the irradiation (via the DED) and the host be taken into account.

## 6. Discussion

This Section discusses (i) the influence of the redox potential on nucleation and growth, and the analogy with photography and (ii) the specific role of the DED in Ag clustering.

**6. 1. Redox dominates nucleation and growth**

Our data show the crucial role of photoelectrons and photo-holes, and their interplay with irradiation defects and with the silver species themselves, in the irradiation-induced nucleation and growth of silver nanocrystals in glasses. This led us to a close comparison with pulsed radiolysis studies of $Ag^+$-containing aqueous solutions, which had shown [12] [63] that the stability of small aggregates strongly depends on charge exchanges and that the free energy of the various (charged and neutral) Ag oligomers could be evaluated and compared to that of the (liquid) host components on a redox scale. Briefly, whereas in vacuum the ionization potential of a small cluster tends to *increase* with average size, the electron donor strength of such clusters actually *decreases* with size in aqueous solutions, due to the effect of the surrounding medium solvation energy: the larger the cluster, the easier it is to solvate a charge from it, and the solvated electron swiftly interacts with the medium to form highly reactive free radicals. Such charge exchange effects clearly also play a major role in nucleation and growth of metal clusters in oxide glasses. As regards growth, after reactions (0 – 11) the following reactions occur during annealing (q and l are respectively the charge and the number of atoms in a cluster) and are responsible for the competition between growth and oxidation:

$$Ag_l^{q+} + Ag^0 \Rightarrow Ag_{l+1}^{q+} \qquad (14)$$

$$Ag_l^{q+} + Ag^+ \Rightarrow Ag_l^{(q+1)+} \qquad (14')$$

$$Ag_l^{q+} + \text{defect } (e^-) \Rightarrow Ag_l^{(q-1)+} \qquad (14'')$$

$$Ag_l^{q+} + \text{defect } (h^+) \Rightarrow Ag_l^{(q+1)+} \qquad (14'''),$$

In standard photography as in aqueous solutions, the image developer would be an electron donor molecule, enhancing growth of all clusters whose redox potential is higher than that of the developer – the sequence being $Ag^0$ and $Ag^+$ clustering followed by developer-induced reduction. Subcritical-sized clusters suffer oxidation until total dissolution [64]; depending on the developer's redox potential, the critical number of silver atoms is between 2 and 5. This critical size bears no direct relation to that derived from classical nucleation theory [65]: in the latter it is the result of a competition between surface and volume energies for comparatively large clusters, whereas here it involves charge exchange effects in the very earliest stage of aggregation (a few atoms) [66]. By analogy with photography, the "latent image" formed by the subcritical clusters is unstable, as opposed to the "image" formed by those above the size threshold, the transition between the two being activated by the redox potential of the "developer".

Whereas photographic development in solution only provides electron donor molecules, the situation is more complex in irradiated glasses. The analogy between classical photography and irradiation of silver-containing glasses may be carried further but here, the "developer" is the glass host *and* its charged species populations. The latters' evolution depends on both electrons (reaction 14'') and holes (reaction 14''') being released by defect traps as the irradiated glass is annealed. The redox conditions in which cluster growth may take place are determined not only by the glass chemistry, but - to a larger degree - by the charge states of the Ag species formed upon the release of electrons and holes from the defect traps. The corresponding states (including the defect traps and Ag oligomers) have well-defined free energies, and a redox scale that includes them may be drawn (**Fig. 16**) to indicate the various possible reactions (e.g. clustering versus dissolution of Ag). It is important here to stress that Ag is a fast diffuser in glass, and the charge exchanges that we are interested in here occur at temperatures well below those at which modifications of the basic glass structure can occur. In view of our findings, it is not exaggerated to state that the behavior of Ag in glass is essentially the same as that of Ag in aqueous solutions, to within the orders-of-magnitude difference of Ag diffusion coefficients in the two hosts. The detailed nature and stability of irradiation-induced defects obviously influences the kinetics (i.e., the temperature dependence) of charge release and charge equilibrium, but their main role is that of charge reservoirs.

The differing redox potentials of the reducing and oxidizing defects account for: (i) the critical size (3-4 atoms) below which clusters are unstable versus oxidation by the surrounding medium, while larger clusters may grow; (ii) the base glass composition-dependence of the reduction efficiency: e.g., a small concentration of Sb or As in photosensitive glasses leads to more reducing conditions; (iii) cluster dissolution under the thermodynamical redox potential of the host when the annealing temperature/time is increased. It also indicates that modifications in irradiation DED affect nucleation and growth, as discussed below.

### 6.2. Why DED impacts on redox behavior

Whereas (see § 5) $\gamma$-irradiation leads to a decrease of the $Ag^0$ concentration upon annealing, the large DED produced by ion irradiation favors a higher local concentration of reducing E' centers, leading to an increase in $Ag^0$ and providing a very efficient path for the formation in glass of stable Ag nanoclusters which are then less susceptible to oxidation by NBHOC centers (Fig. 16). As shown in § 4, ESR and OA spectra revealed notable differences between the $\gamma$- and ion-irradiated $Ag^+$-doped glass samples. The ESR spectrum in Fig. 3a, corresponding to a room-temperature γ-irradiated glass, displays an intense contribution from the hole-trapping NBOHC defects whereas its counterpart obtained from room-temperature, 1.6 MeV He or 12 MeV Br ion-irradiated glasses (**Fig. 3b**) contains approximately equal contributions from both the NBOHC and electron-trapping E' centers. *This DED-induced difference in the surviving population of electron- and hole-trapping defects provides a major source for the difference in the nanocluster formation probability.* As noted above, OA and ESR measurements showed that the NBOHC density was essentially the same in both cases. Due to network modifiers such as Na or Ca, silicate glasses contain high concentrations of Non Bridging Oxygen (NBO), and these act as preferential hole traps to form NBOHC, the center that dominates the ESR spectrum of **Fig. 3a**. On the other hand, the enhanced formation of oxygen Frenkel pairs in the ion-irradiated (high DED) case presumably accounts for significant electron trapping in E' centers, an effect that is very weak under the low energy transfers involved in γ-irradiation [46], [67], [68]. It is difficult to go into more detail. The difference between the relative amplitudes of the electron trap and NBOHC signals in the ESR and OA spectra of the γ-irradiated glass is at least partially due to the contribution of multivalent Ag oligomers to a charge imbalance in the glass (compare the OA spectra before and after the introduction of Ag in the glass). Knowledge of electron traps is rather poor: although the existence of the electron trap identified in our OA spectra is known, its structure has not been established. Griscom [69] suggested that electrons could be trapped on alkali clusters in glasses. As noted above, another contribution (which was not studied here) is probably the dependence of the optical absorption on the medium polarity and the interaction of reduced Ag with neighbors.

Although more work on defect identification remains to be done, the main result of the comparison (**Fig. 3**) is that, in spite of a relatively weaker reducing efficiency than that of γ irradiation, ion-irradiation and its correspondingly large DED is far more effective in producing metal nanocluster precipitation. The high charge density in the ion tracks has two opposite consequences: it enhances electron-hole recombinations, thus reducing the clustering efficiency, but on the other hand, the large residual electron density leads to rapid silver cluster nucleation which is less susceptible to oxidation during annealing. The lower efficiency being compensated by a much higher total ion fluence (versus the $\gamma$ dose), the final reduced silver concentration is significantly higher.

### 6. 3. A "phase diagram" for Ag nanocluster nucleation and growth

In order to summarize our results and their consequences, we propose a compound, unifying picture of all the nanocluster nucleation and growth processes in glasses. It is clear, from the discussion above, that the different processes are not on an equal footing. Those that are determined by glass chemistry can be described by an equilibrium phase diagram in the (T, redox) plane. Those that involve irradiation-induced charged species are determined by nonequilibrium effects such as carrier concentrations, recombination paths, etc. We present (**Fig. 17**) a qualitative three-dimensional "phase diagram", where the quotation marks indicate this extension from equilibrium to nonequilibrium processes. In addition, of course, there will be kinetic effects due to carrier lifetimes, thermally activated defect trapping and detrapping, etc.

Two of the variables in Fig. 17 are familiar: the temperature, and the quantity termed "glass acidity" by glass-makers which qualifies the initial (equilibrium) redox state $E_0$ of the host, depending on the nature and the amount of electron donor it contains. The annealing duration (30 min in the example of Fig. 17) is assumed to be constant throughout. The results deduced from standard glass chemistry - no defects, no irradiation - correspond to those drawn in the (T, redox) plane (it typically portrays the compositional change between silica and silicate glasses, as well as the efficiency of adding such reducers as Sb or As in order to scavenge holes and thus enhance metallic nanocluster formation). The novelty is in the addition of a third variable: the ionizing radiation's deposited energy density (DED). The concept of DED has the advantage that it is an easily defined and measured quantity related to the irradiating beam, depending only on the target's elements, mass and density, and independent of its thermodynamical or chemical state. It does not assume anything about the processes occurring when the energy is deposited in the target's electron bath. From the point of view adopted here, the DED provides a measure of the *initial charge (electron and hole) density* injected into the glass by irradiation. As such, it is independent of the other variables. It is not a thermodynamic variable, hence the quotation marks above. As seen previously, the initial DED influences nucleation and growth indirectly, via both primary charge injection and defect creation, but - because Ag is such a fast diffuser - defects can be treated simply as charge reservoirs. Thus the charge balance is modified via defect kinetics, as shown notably in the (DED, T) plane. The charge balance (and resulting aggregation probability) also depend on the combined influence of the DED and average host redox potential, as shown notably in the (DED, $E_0$) plane. The "phase diagram" drawn here is heuristic rather than quantitative, but in spite of this limitation, it provides both a summary of our results and a guide for nucleation and growth control.

There is a low-valued region of the "phase diagram" in which $Ag^+$ remains simply dissolved in the glass. As the temperature is raised, there is first a rather broad region in which an increase in the DED and/or (separately or – more efficiently – simultaneously) in the reducing efficiency leads to *both* nucleation (with or without charge capture) and redissolution. Upon further raising the temperature so that monomer mobility sets in, and for large values of the DED or of the reducing efficiency, significant nucleation and growth occur (involving neutral and/or charged Ag oligomers, as discussed above). At sufficiently high temperatures, dissociation occurs and $Ag^O$ is again dissolved. The transitions between the different regions of the diagram illustrating the Ag population's evolution are, of course, progressive rather than sharp. As portrayed by the asymmetryin the diagram, the DED is very effective in biasing the system towards nanocluster nucleation. A most useful feature of the diagram is to emphasize (see the two examples given in **Fig. 18)** that trajectories exist in the forward quadrant along which combined control over both DED and glass acidity may lead to enhanced nanocluster tailoring [70] by controlling the nucleation and growth speeds.

## 7. Conclusion

We have studied the initial stages of Ag clustering in glasses, as well as the stages of further growth and the clustering efficiency, under irradiation at differing deposited energy densities (γ versus ion irradiation). Our main results are as follows. (1) Initial aggregation as well as latter-stage cluster growth depend simultaneously on the charge state of the Ag species and on the host redox potential; (2) The latter depends, in turn, on the (temperature-dependent) release of charges by the different electron- and hole-traps produced by the γ or ion energy depositions; (3) Such irradiations (coupled to appropriate annealing) are an efficient means of modifying and controlling the system's redox state via both charge evaporation from defect reservoirs and transitions between Ag oligomer charge states; (4) The corresponding modifications supplement and interact with those due to the chemical composition of the host itself; (5) The large difference (a factor ~ $10^4$) between energy densities deposited by γ versus ion irradiation has striking consequences (see § 5) on the oligomer production efficiency and on the final cluster density because of the different DED morphologies in the two cases. As compared to the effect of $\gamma$ irradiation, the higher overall DED of the ion irradiation actually leads to a far lower initial $Ag^0$ production efficiency because of charge recombinations occuring in the plasma along the ion path. However, ultrafast clustering of neutral Ag atoms also occurs in this plasma; combined with the high total ion fluence, this leads to a significantly higher final density of stable Ag nanoclusters.

A strong analogy between our results and those previously found in aqueous solutions is apparent, an interesting feature being that early intermediate states of charge evolution and aggregation that were difficult to observe in aqueous solutions are visible in glasses because of the slower Ag and charge diffusion.

In addition to the usual surface and volume energies, a correct description of nucleation and growth of metallic nanoclusters in glass (and probably other polar media) should thus take into account the redox (charge) state of clusters. Cluster growth is governed not only by $Ag^0$ exchange but also by charge (via defects) and $Ag^+$ exchange (reactions 14). There is a competition between fast reduction, favored at high DED, and corrosion by hole defects – the latter being ineffective when the cluster size exceeds a critical nuclearity threshold. Quite generally, in insulators (and semiconductors) – and notably in glasses – defects and chemical species may have more than one charge state; Reaching equilibrium between charge states is often a long-term process. This is the source of the influence of redox effects on nucleation and growth. Combining the classical nucleation scheme with the charge recombination mechanisms described by Equations 14 could possibly be treated in the generalized approach to nucleation suggested by Binder and Stauffer [71]. It is a task well beyond the scope of the present work, but we argue that these effects cannot be ignored when considering nanocluster nucleation and growth in such materials.

## Acknowledgments

We are grateful to F. Jomard (LPSC-CNRS), for his contribution to the SIMS experiments and to M. Lourseau (LPC-CNRS), C. Boukari and O. Kaitasov (CSNSM-CNRS) for assistance with irradiations, as well as to B. Boizot and S. Esnouf (LSI-Palaiseau) for their contribution to the implementation and interpretation of the ESR experiments. We thank M. Jean, H. Cocquard, E. Gesell, A. Godin, T. Montigny, D. Brochot (Corning Europe Research) for technical assistance. Discussions with A. Broniatowski are also gratefully acknowledged.

***[ Notes added***: After completion of this paper, we became aware of

1. the paper by Shihai Chen, Tomoko Akai, Kohei Kadono and Tetsuo Yazawa, *Appl. Phys. Lett.* ***79****, 3687 (2001)* entitled: "Reversible control of silver nanoparticle generation and dissolution in sodalime silicate glass through x-ray irradiation and heat treatment", showing how alternating high-temperature anneals and x-ray irradiations can successively dissolve and re-form Ag nanoclusters. This allows a measure of size control. Albeit limited to photon irradiation effects, the authors' interpretation of their results in terms of the interaction of Ag with electron- and hole-center defects is rather similar to that presented here.

2. a preprint entitled "Formation mechanisms of radiation-induced defects in high-purity silica glass" by J. Roiz, J. C. Cheang-Wong, L. Rodríguez-Fernández, E. Muñoz, R. Ortega-Martínez, A. Crespo, J. Rickards and A. Oliver, which describes defect formation and identification under conditions (gamma-, electron- and ion-irradiation) that are quite relevant to the present work. We are grateful to Prof. A. Oliver for comunicating this paper prior to publication.]

**Table 1**

Summary of experiments performed and information obtained in this work.

| samples | irradiation (irrad. temp.) | anneals (K) | experiments | identification of |
|---|---|---|---|---|
| **Silicate glass (no Ag)** | $^{60}Co$ $\gamma$ photons ($T_{irr}$ = 80 K, 300K) | 80 - 843 | ESR<br>OA | defects |
| **Silicate glass + Ag (65 appm)** | unirradiated | 300 - 843 | OA | absence of nanoclustering |
| **Silicate glass + Ag (17-130 appm)** | $^{60}Co$ $\gamma$ photons ($T_{irr}$ = 80 K, 300 K) | 80 - 802 | ESR<br>OA | defects, Ag oligomers<br>defects,Ag oligomers, nanocluster formation |
| **Silicate glass + Ag (65-130 appm)** | He (1.6 MeV), Br (12 MeV) ($T_{irr}$ = 120 K, 300 K) | 300 - 693 | ESR<br>OA | defects<br>nanocluster formation |
| **Silicate glass + implanted Ag (500 appm)** | unirradiated<br>------------<br>Br (12 MeV) ($T_{irr}$ = 300 K) | 300 - 823 | SIMS<br>OA<br>-----------<br>OA | diffusion profile<br>no clustering<br>--------------<br>influence of DED on nucleation |
| **Silica + implanted Ag (500 appm)** | unirradiated<br>Br (12 MeV) ($T_{irr}$ = 300 K) | 300 - 1073 | OA | influence of host redox vs. DED on clustering |

**Table 2**

Summary of irradiating conditions. Gamma irradiation parameters are given in terms of dose, dose rate and linear energy transfer (LET). The corresponding terms for ion irradiations are fluence, ion flux and deposited energy density (DED).

| Radiation (ion beam) | $\gamma$ | **$Ag^+$** | **$Br^{7+}$** | **$He^+$** |
|---|---|---|---|---|
| Energy | 1.17-1.33 MeV | 480 keV | 11.9 MeV | 1.6 MeV |
| Dose rate (ion flux) | 3 kGy $h^{-1}$ | $2x10^{12}$ (Ag $cm^{-2}$ $s^{-1}$) | $(4 - 9)x10^{10}$ (ions $cm^{-2}$ $s^{-1}$) | $10^{12}$ - $10^{13}$ (ions $cm^{-2}$ $s^{-1}$) |
| Dose (fluence) | 21 - 40 kGy | $10^{15}$ ions $cm^{-2}$ | $10^{11}$-$10^{14}$ ions $cm^{-2}$ (*) | $\leq 10^{15}$ ions $cm^{-2}$ |
| (DED) LET | 0.23 eV $nm^{-1}$ | 0.4 $keVnm^{-1}$ | 4 $keVnm^{-1}$ (2 – 2500 $eVnm^{-3}$) | 0.3 $keVnm^{-1}$ |
| dose rate ($eVnm^{-3}$ s) | $10^{-5}$ | 70 | 1-2 | 4-40 |
| Penetration depth | >> 1 cm | ~ 0.15 µm | ~ 4.8 µm | ~ 5.2 µm |

* in terms of energy absorption, this corresponds to $(2\times10^2 - 2\times10^5)$ kGy

## Table 3

Adjustment parameters (assuming Gaussian components) of the OA spectra (Fig. 6) from an undoped and a 17 appm doped silicate glass after room temperature gamma irradiation (40 kGy). The position and width of the first four absorption bands are determined from the undoped glass, and assumed to be unchanged when determining the neutral silver band position (at 3.6 eV).

| Position (eV) | Width (eV) | Intensity ($cm^{-1}$) | | Assignment |
|---|---|---|---|---|
| | | undoped | doped | |
| 1.98 | 0.3 | 1.0 | 0.19 | NBOHC |
| 2.79 | 0.65 | 2.1 | 0.41 | NBOHC |
| 4.1 | 1 | 2.45 | 0.86 | $e^-$ |
| 5.35 | 0.65 | 4.1 | 1.7 | POR |
| 3.6 | 0.5 | - | 0.5 | $Ag^0$ |

# Figures

**Fig. 1**: Optical absorption spectra of the unirradiated silicate glass chemically doped with 65 appm silver per host atom. Annealing temperatures and durations as shown (in this experiment the hump at 600 nm was due to a grating artefact, and should be disregarded). No absorption band due to reduced silver is apparent. Sample thickness: 1 mm. (Vertical axis units multiplied by 100).

**Fig. 2**: SIMS concentration profile of $^{107}$Ag in silicate glass after implantation at 480 keV (fluence $10^{15}$ ions.cm$^{-2}$), followed by annealing at increasing temperatures. The silicate glass atomic density is 6 x $10^{22}$ at.cm$^{-3}$.

**Fig. 3**: ESR spectra of chemically doped (65 appm Ag) silicate glass irradiated by (a) gamma photons from a $^{60}$Co source (15 kGy); (b) Br (12 MeV, $10^{14}$ at.cm$^{-2}$, full line) and He (1.6 MeV, $10^{15}$ at.cm$^{-2}$, dashed line) ions. Irradiations were performed at 120 K and spectra were taken on unannealed samples. Inset in spectrum (a) shows the weak signal associated with a trapped electron.

**Fig. 4**: ESR spectra of chemically doped (65appm Ag) silicate glass after irradiation by 1.6 MeV He ions ($10^{15}$ cm$^{-2}$), annealed 1 hr. at increasing temperatures. The curves are shifted for clarity.

**Fig. 5**: ESR signal of chemically doped (65 appm Ag) silicate glass, gamma irradiated (15 kGy) at 77 K and annealed as shown. See text for identification of the different paramagnetic

Ag centers. For clarity, the high intensity signal around H=3350G, due to irradiation defects, has not been drawn, and the curves have been shifted.

**Fig. 6**: Room temperature OA spectrum and its component analysis for a chemically doped (17 appm Ag) silicate glass, obtained after a 40 kGy room-temperature gamma irradiation. The component parameters and their assignments are given in **Table 2**. The spectrum of the unirradiated glass does not contain the 3.6 eV band due to $Ag^0$.

**Fig. 7**: Room temperature OA spectra of chemically doped (118 appm Ag) silicate glass after a 21 kGy gamma irradiation, and post-annealing for 30 min. at the temperatures shown. Sample thickness : 1 mm.

**Fig. 8**: Integrated ESR signals (proportional to their concentration) for different paramagnetic centers, as a function of annealing temperature (annealing time 30 min).

**Fig. 9**: Room temperature gamma irradiation dose-dependence of the maximum optical absorbance values for different bands, resulting from multigaussian decomposition. The bands correspond to different defects and to $Ag^0$, see text and Fig. 6. This result was obtained for a chemically doped (118 appm Ag) silicate glass. Lines are Poisson curve adjustments with n=1 (see Eq. 12 and text).

**Fig. 10**: Optical absorption of chemically doped (118 appm Ag) silicate glass, gamma-irradiated to a dose of 21 kGy at room temperature, and annealed for 30 min. at the temperatures shown.

**Fig. 11**: Long-term dependence of room temperature (RT) OA spectra for two identical chemically doped (118 appm Ag) silicate glass samples, gamma-irradiated to a dose of 21 kGy and annealed at (a) 100 °C for 30 min or (b) 140 °C for 30 min.

**Fig. 12**: OA of chemically doped (130 appm Ag) silicate glass, irradiated with Br ions (12MeV, $10^{14}$ at.cm$^{-2}$), and post-annealed at temperatures shown. Sample thickness : 1 mm. (Vertical axis units multiplied by 100).

**Fig. 13**: $Br^{7+}$ (12MeV) ion irradiation fluence-dependence of neutralized silver quantity in silicate glass after annealing at 643 K, as deduced from two separate OA experiments. The saturation concentration corresponds to approx. $4 \times 10^{15}$ Ag cm$^{-2}$. We conclude that at least 75% of the Ag in the irradiation depth was clustered.

**Fig. 14**: OA spectra of Ag-implanted (480 keV, $10^{15}$ ions.cm$^{-2}$) pure silica after annealing as indicated. Upper: Evolution upon annealing only. The irradiation defect contribution to the spectra disappeared almost completely at 573 K. Absorption peaks at $\lambda$<400nm in the Ag-doped silica OA spectra are ascribed to a mixture of silver oligomers (see text), evidencing the first stages of growth. The peak around $\lambda$ = 400 nm is the silver cluster SPR. Lower: Evolution after 12 MeV Br irradiation at a fluence of $2 \times 10^{14}$ ions cm$^{-2}$, followed by annealing as shown. The absorption at $\lambda < 400$ nm due to silver oligomers no longer appeared; we only observed the SPR of Ag clusters at 400 nm. The curves are shifted for clarity. (Vertical axis units multiplied by 100).

**Fig. 15**: OA spectrum of Ag-implanted ($10^{15}$ ions.cm$^{-2}$; 480 keV) silicate glass after by Br ion ($10^{14}$ ions.cm$^{-2}$; 11,9 MeV) post-irradiation and annealing as indicated. As opposed to the silica-based sample (Fig 14), the Br ion irradiation produced an OA signature of $Ag^0$ ($\lambda$ = 360 nm) in the unannealed sample. Annealing produced a progressive increase of the nanocluster SPR signature (400 nm) up to 620 K, followed by a decrease (see text for discussion). For clarity, the curves are shifted by a constant amount of 2 x$10^{-2}$. (Vertical axis units multiplied by 100).

**Fig. 16**: Redox potential ($E^0$) diagram of species observed in our irradiated glasses.

**Fig. 17**: Schematic "phase diagram" describing Ag oligomer and nanocluster evolution in terms of temperature, ion irradiation DED and redox potential of the base host (see text).

**Fig. 18**: Cuts along two different directions of the "phase diagram" shown in Fig.15. (1) lhs: the cross-section parallel to the basal plane illustrates how clustering may be due to (a) the reducing character (redox potential $E^0_{glass}$ ) of the glass composition – leading to equilibrium thermodynamical precipitation – or to (b) the reducing power, via charge neuralization, of the ionizing radiation's DED (nonequilibrium precipitation), or to (c) a combination of these features that both modify redox precipitation conditions. (2) rhs: a planar cross-section containing the temperature axis illustrates the precipitation temperature-dependence in a situation where both the glass composition and the ionizing radiation DED determine the effective redox potential $\langle E^0_{sys} \rangle$.

Fig.1

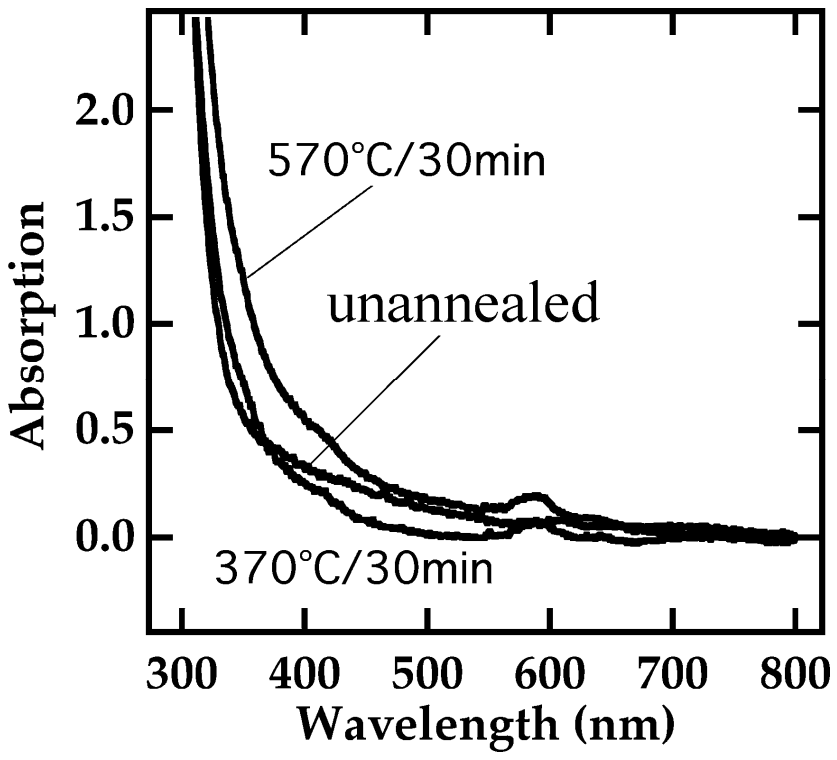

Fig.2

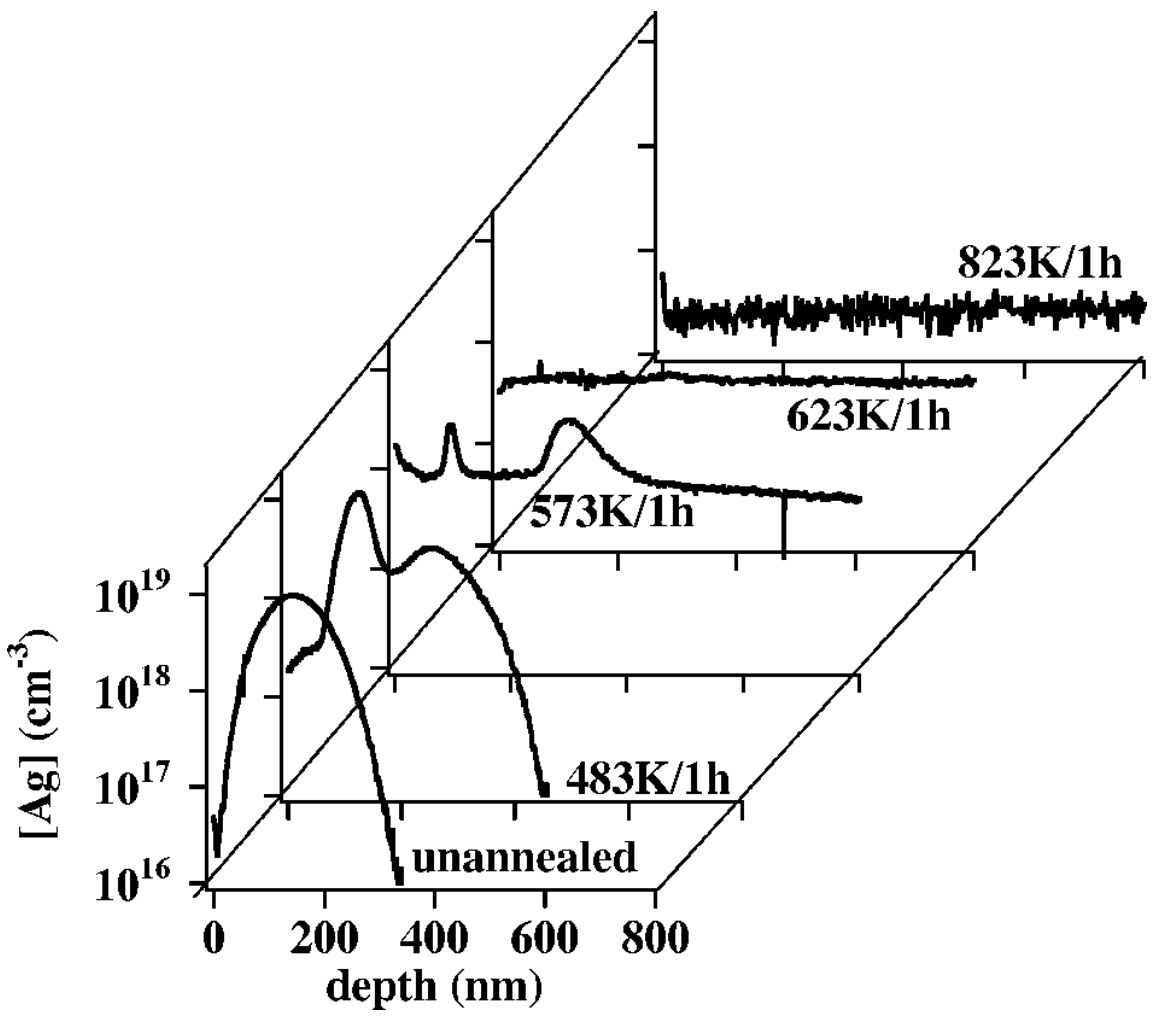

Fig.3

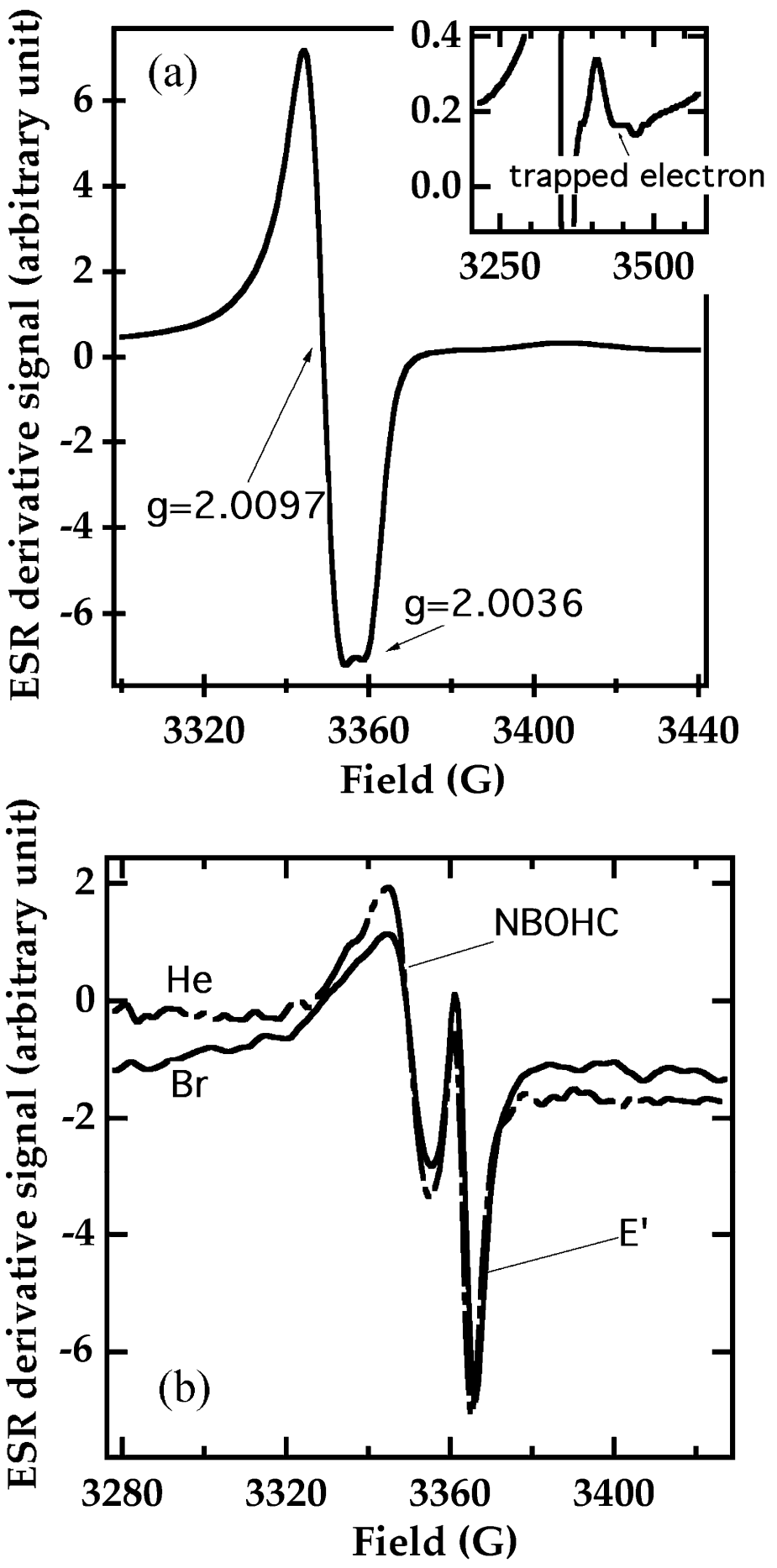

Fig.4

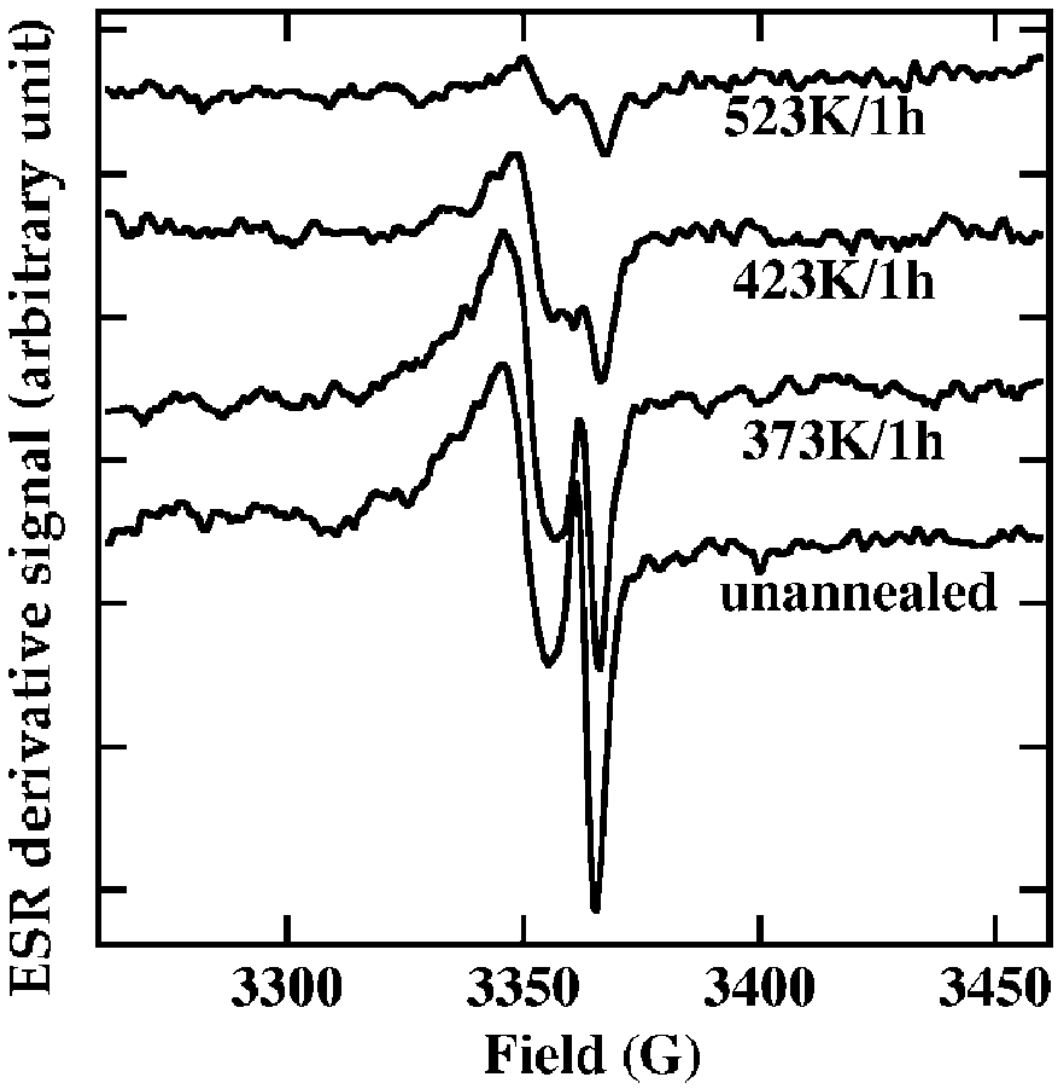

Fig. 5

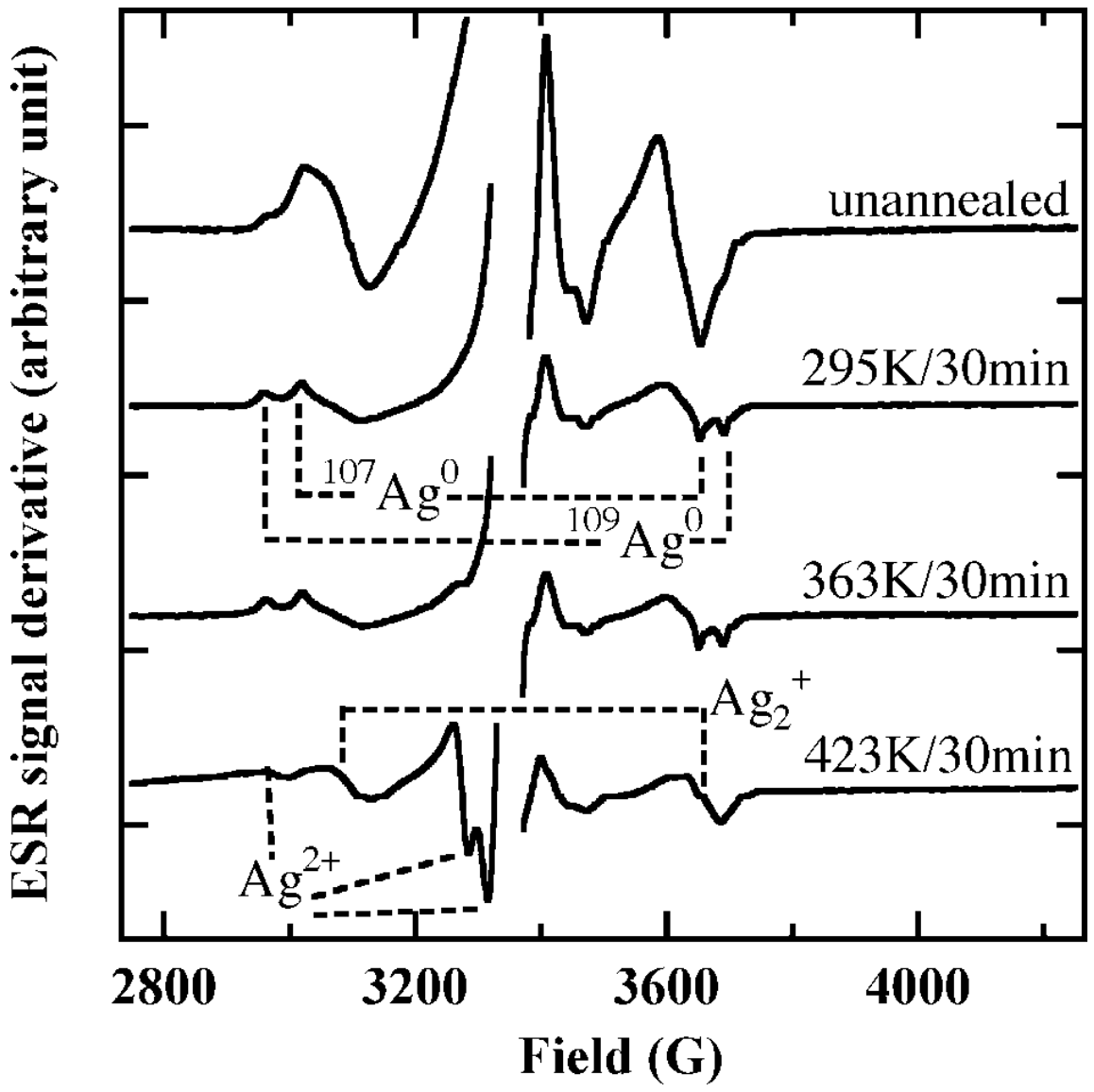

Fig.6

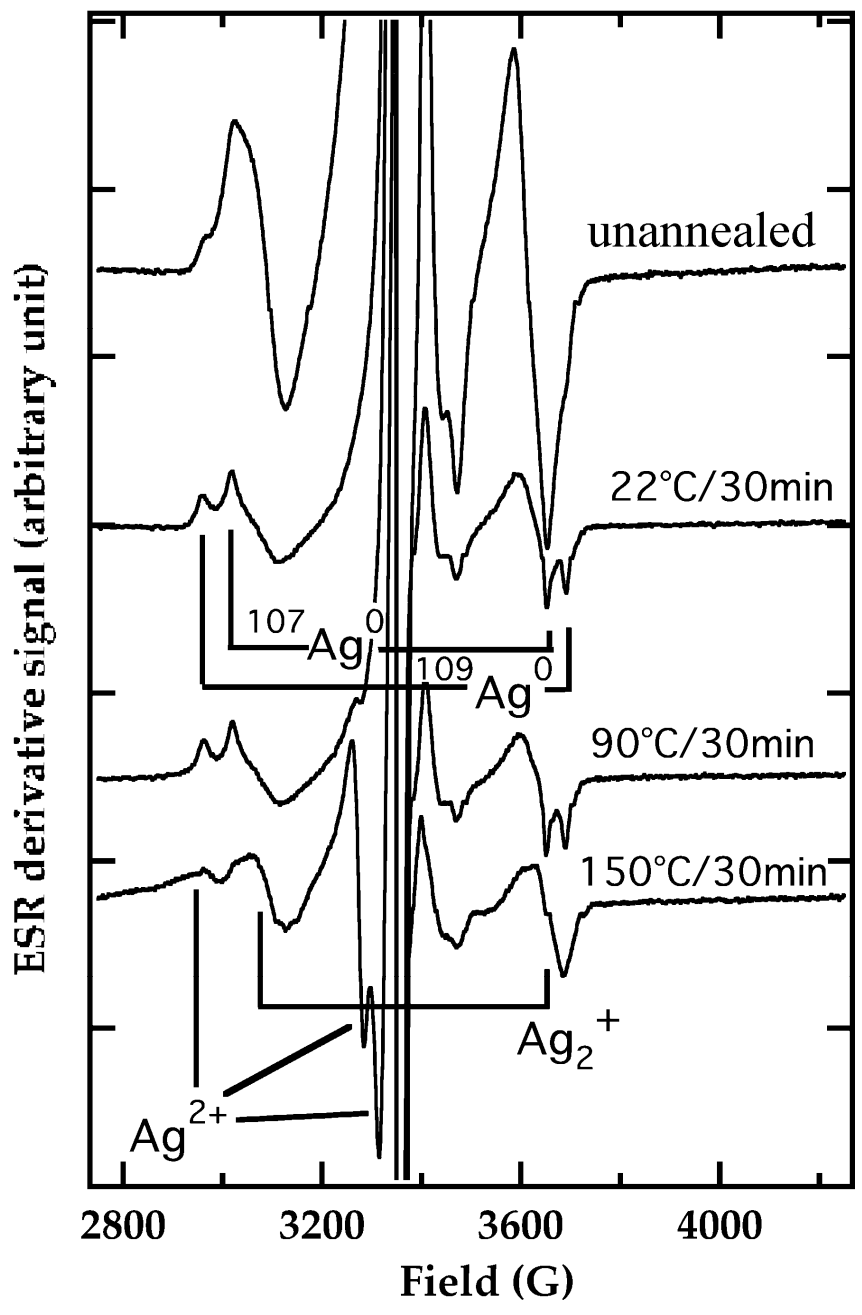

Fig.7

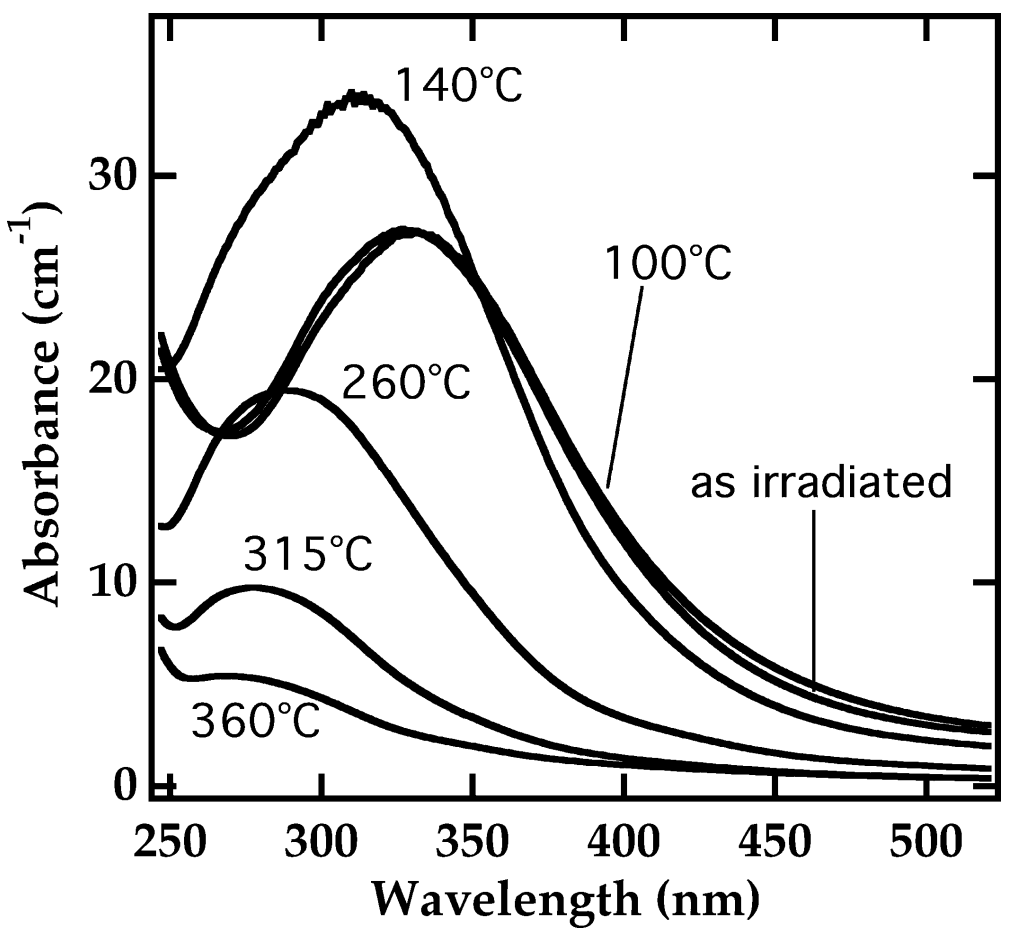

Fig.8

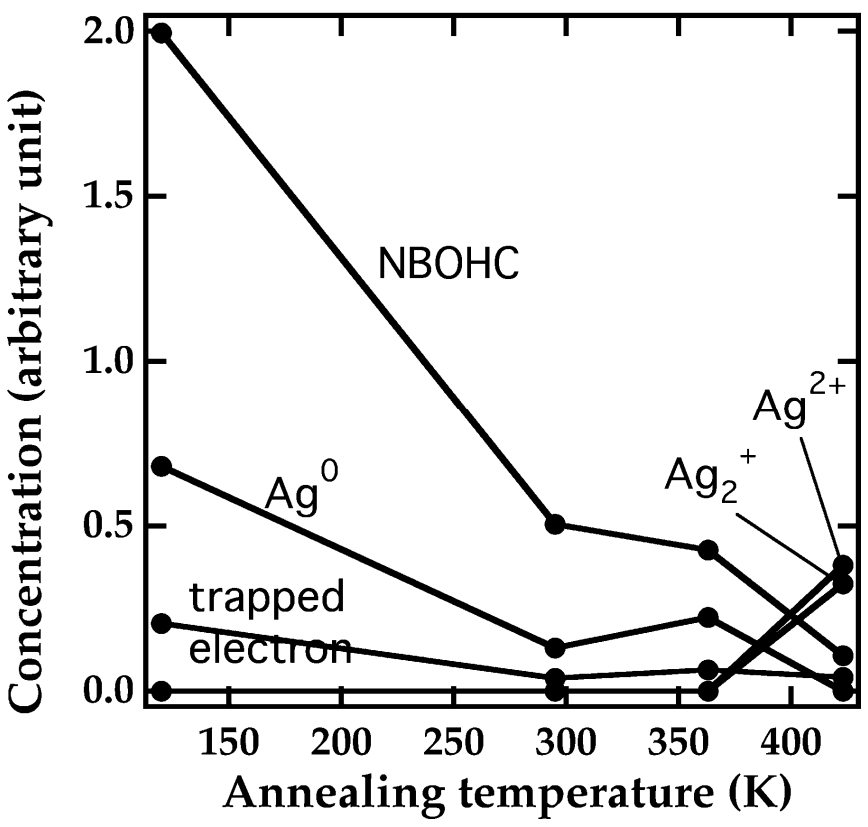

Fig.9

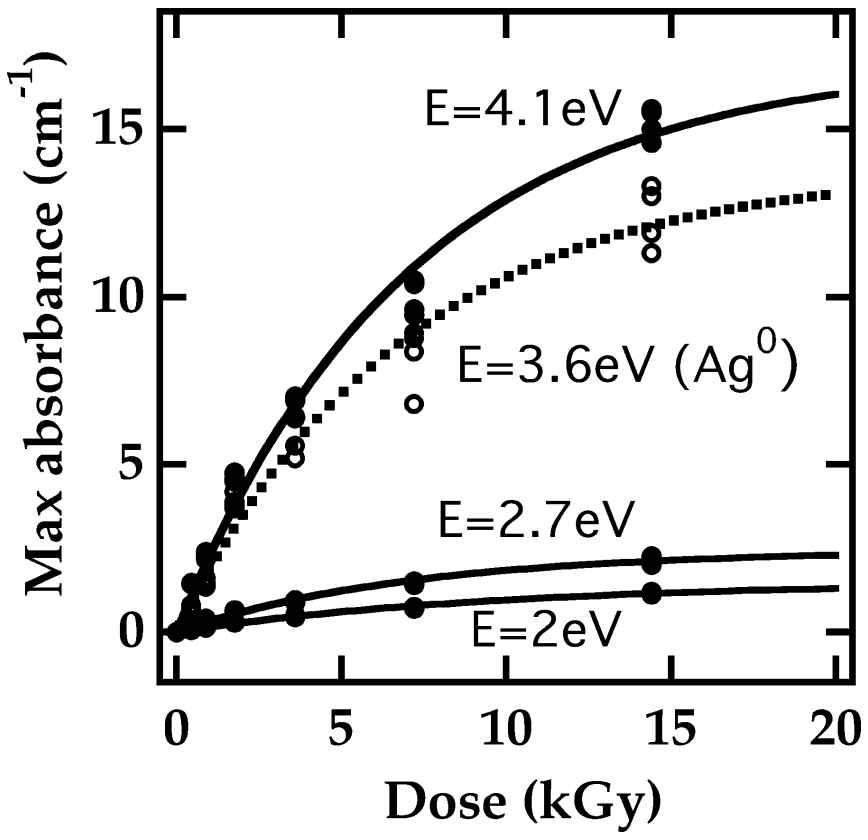

Fig.10

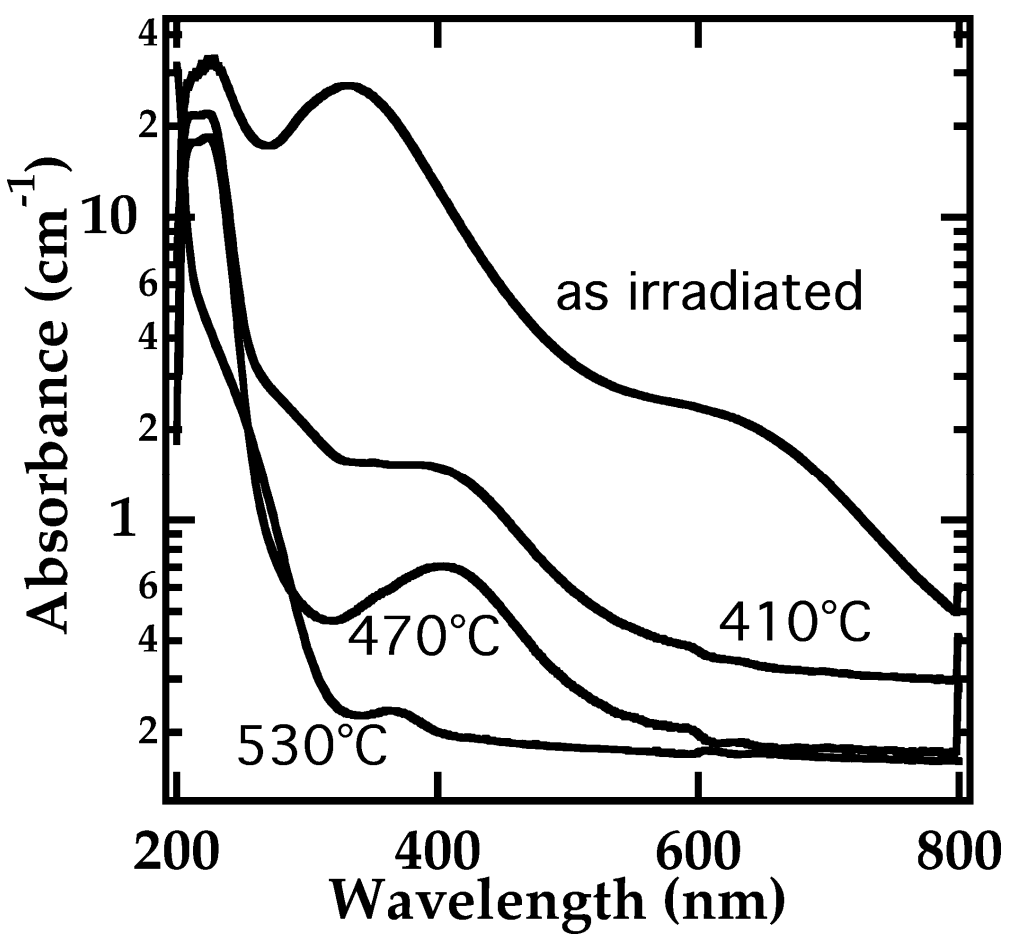

Fig.11

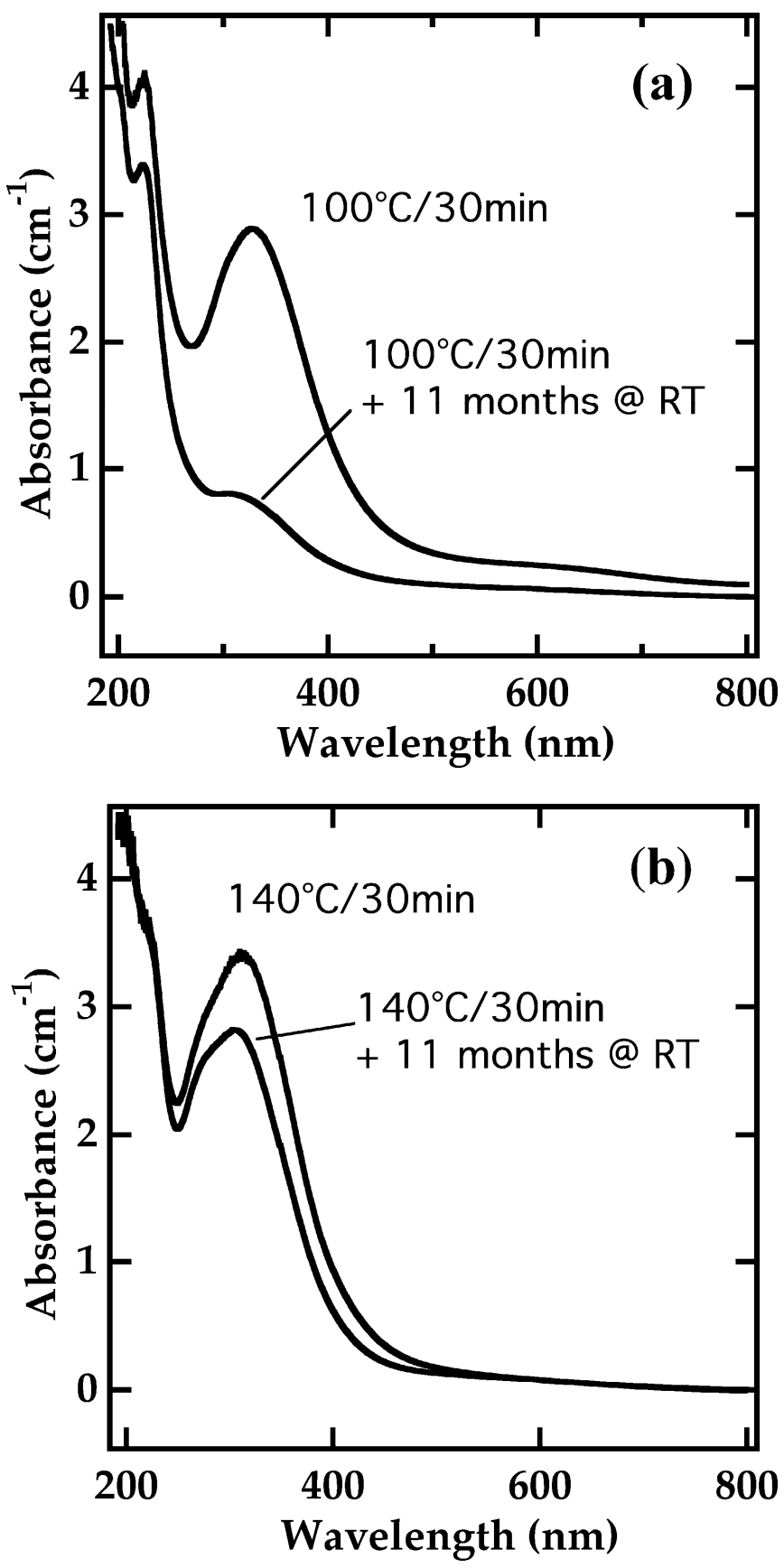

Fig.12

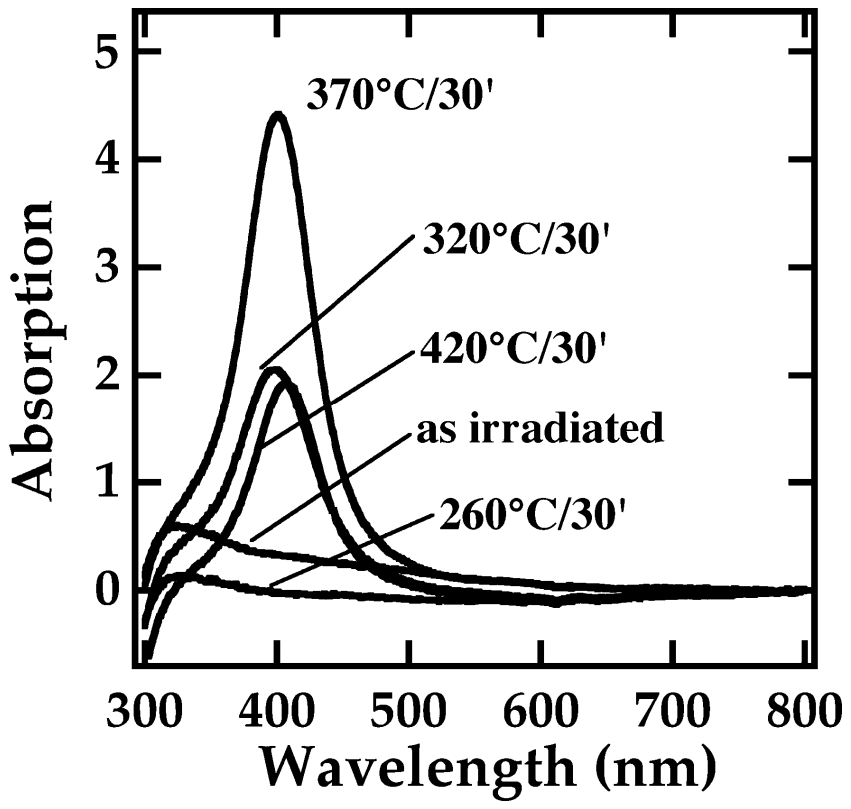

Fig.13

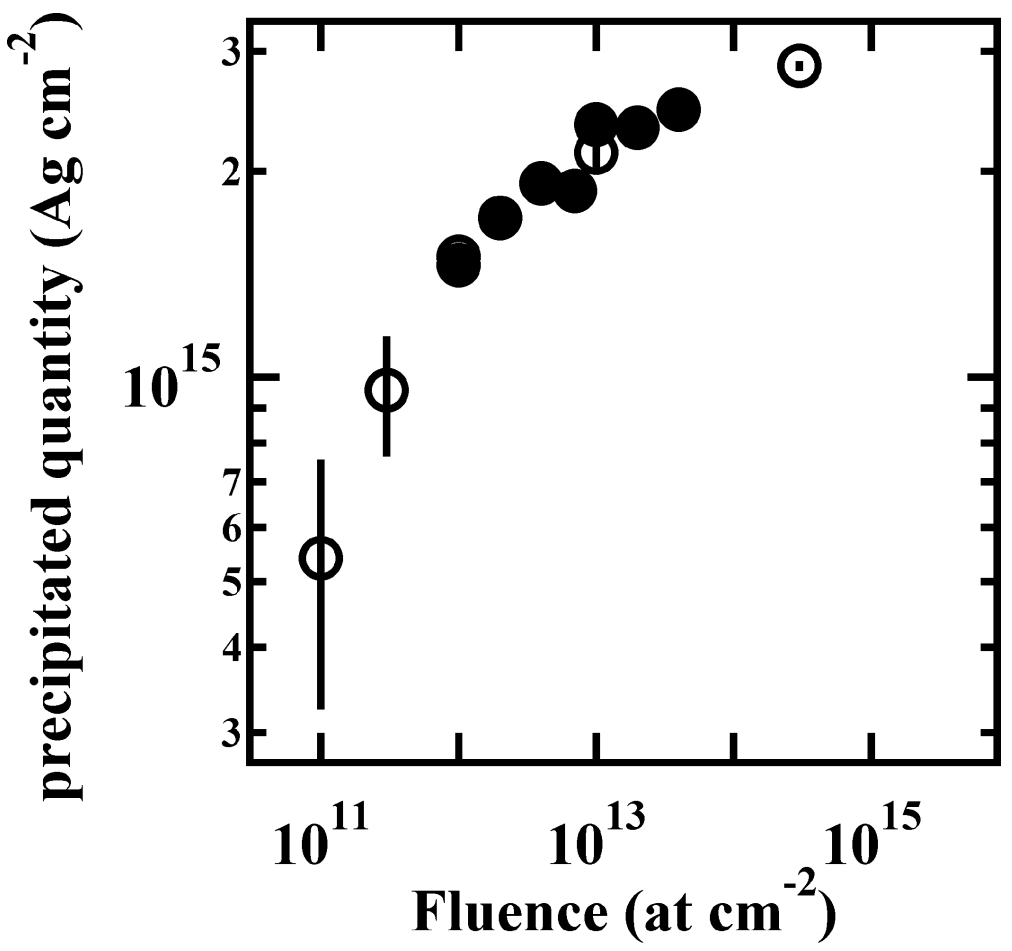

Fig.14

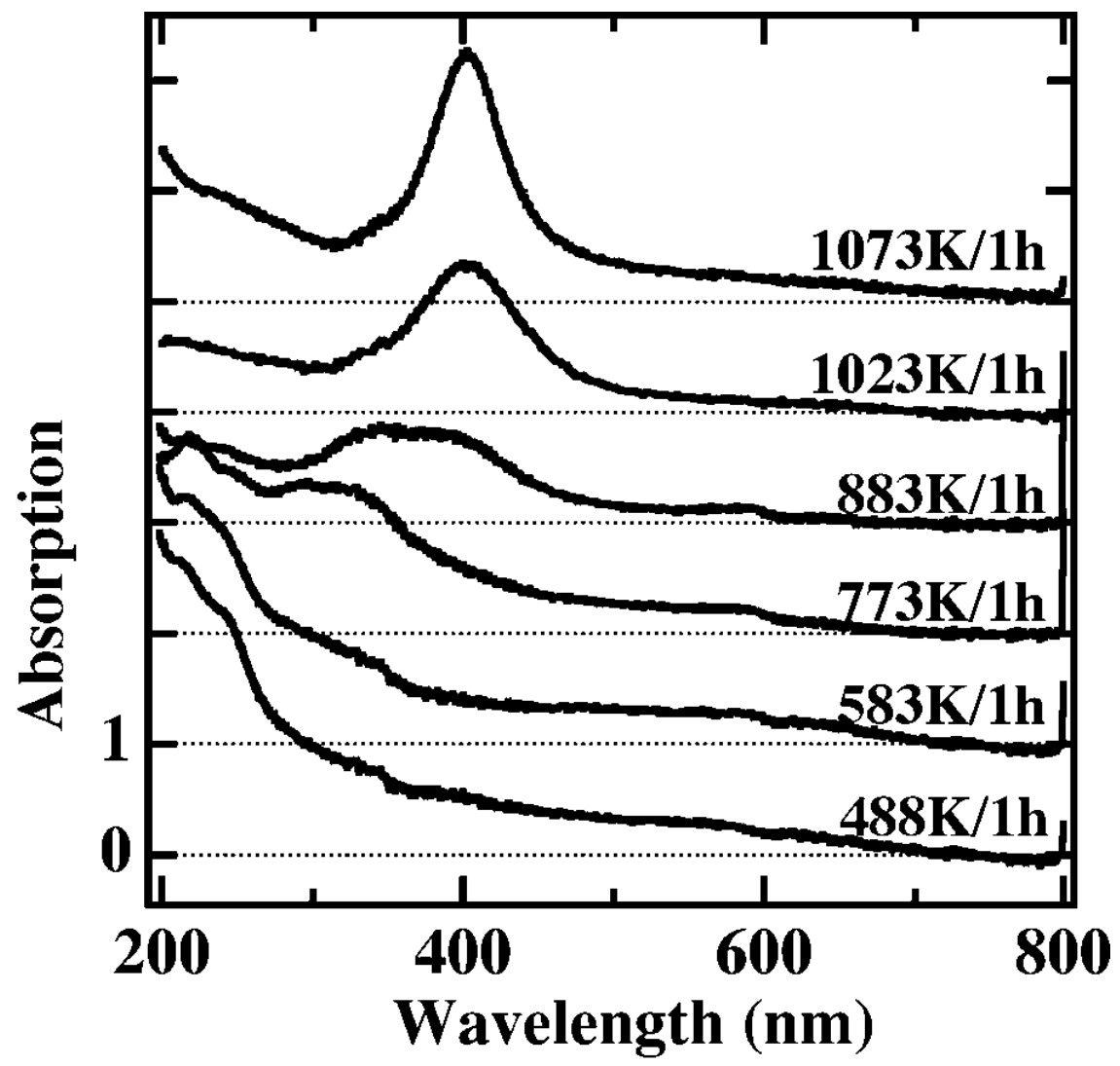

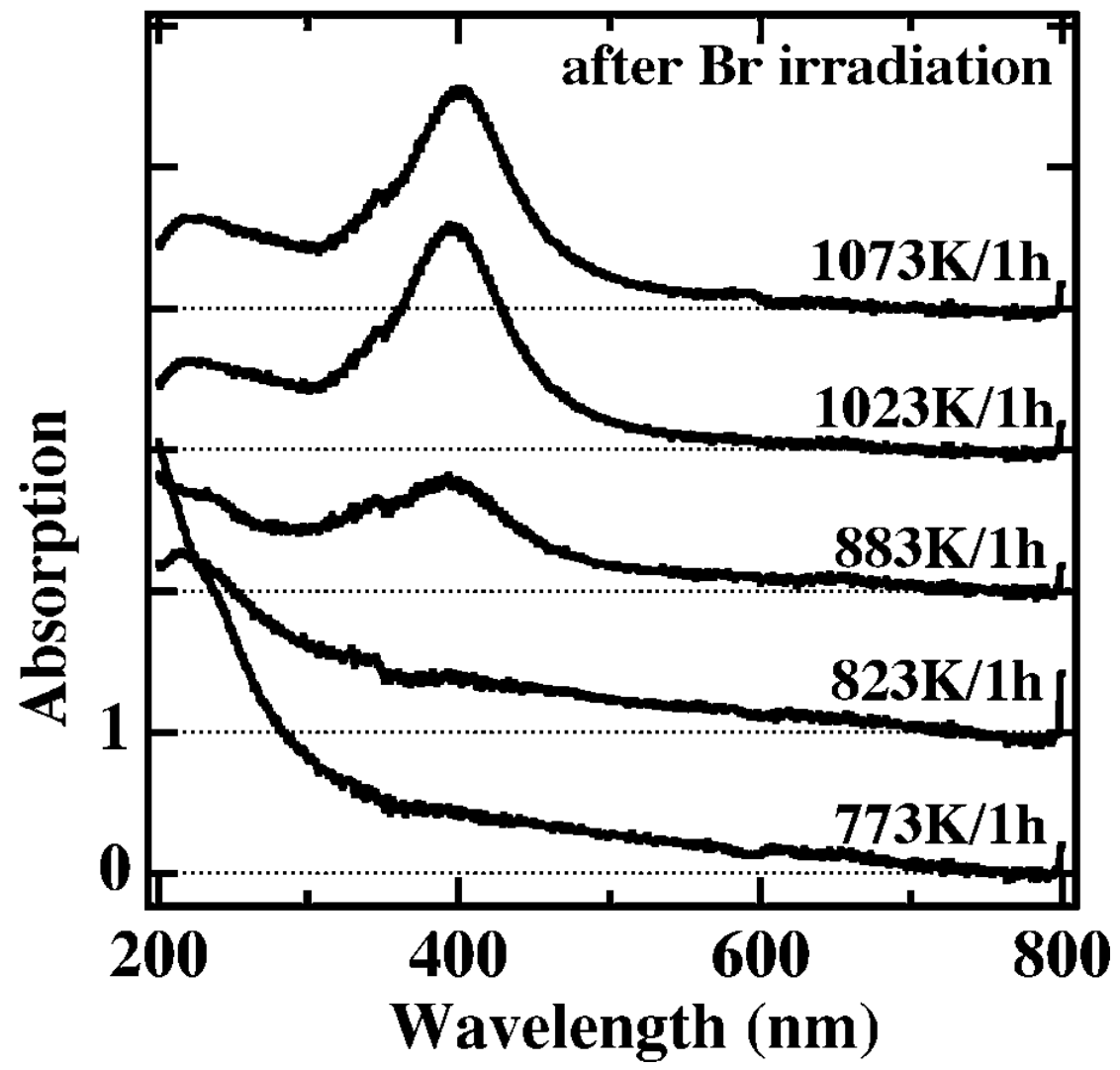

Fig.15

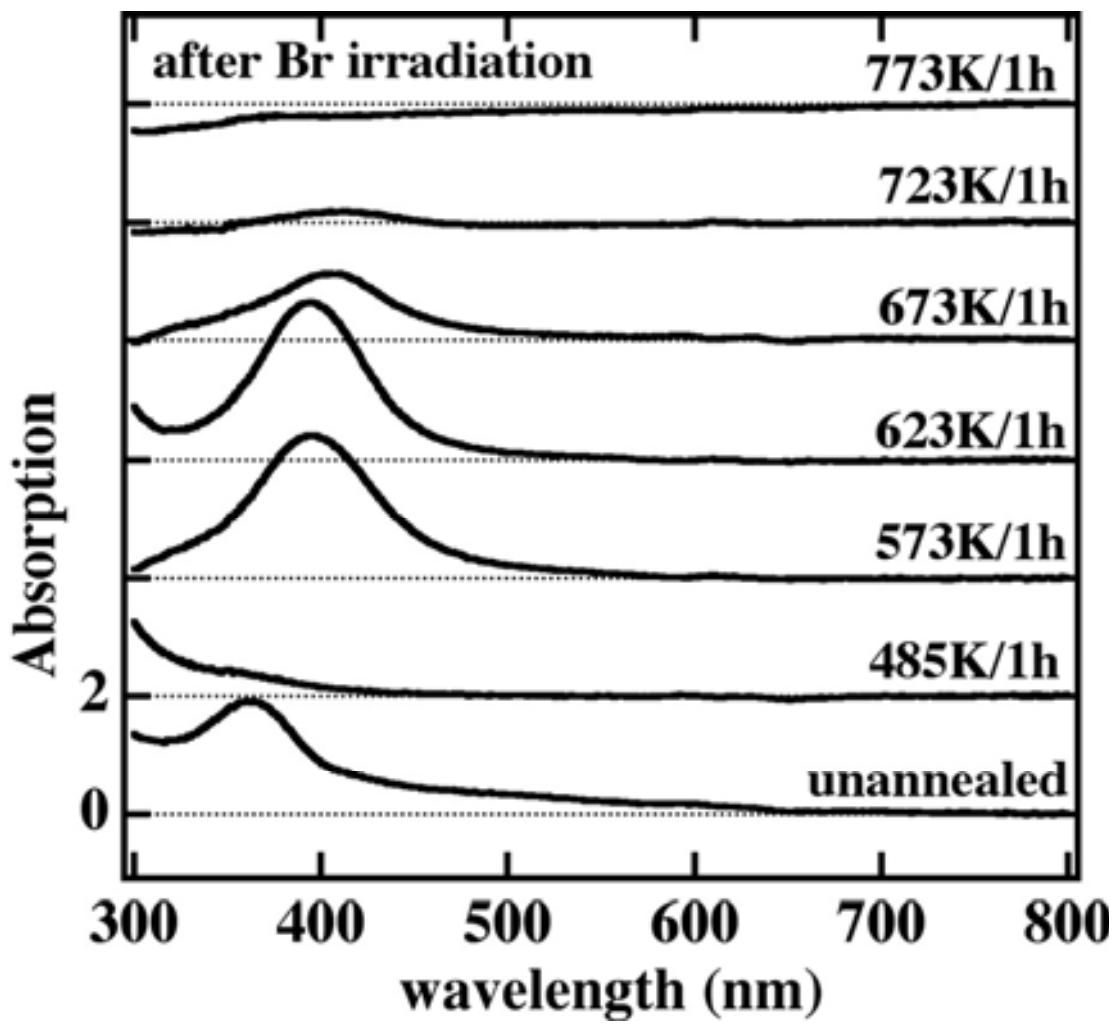

Fig. 16

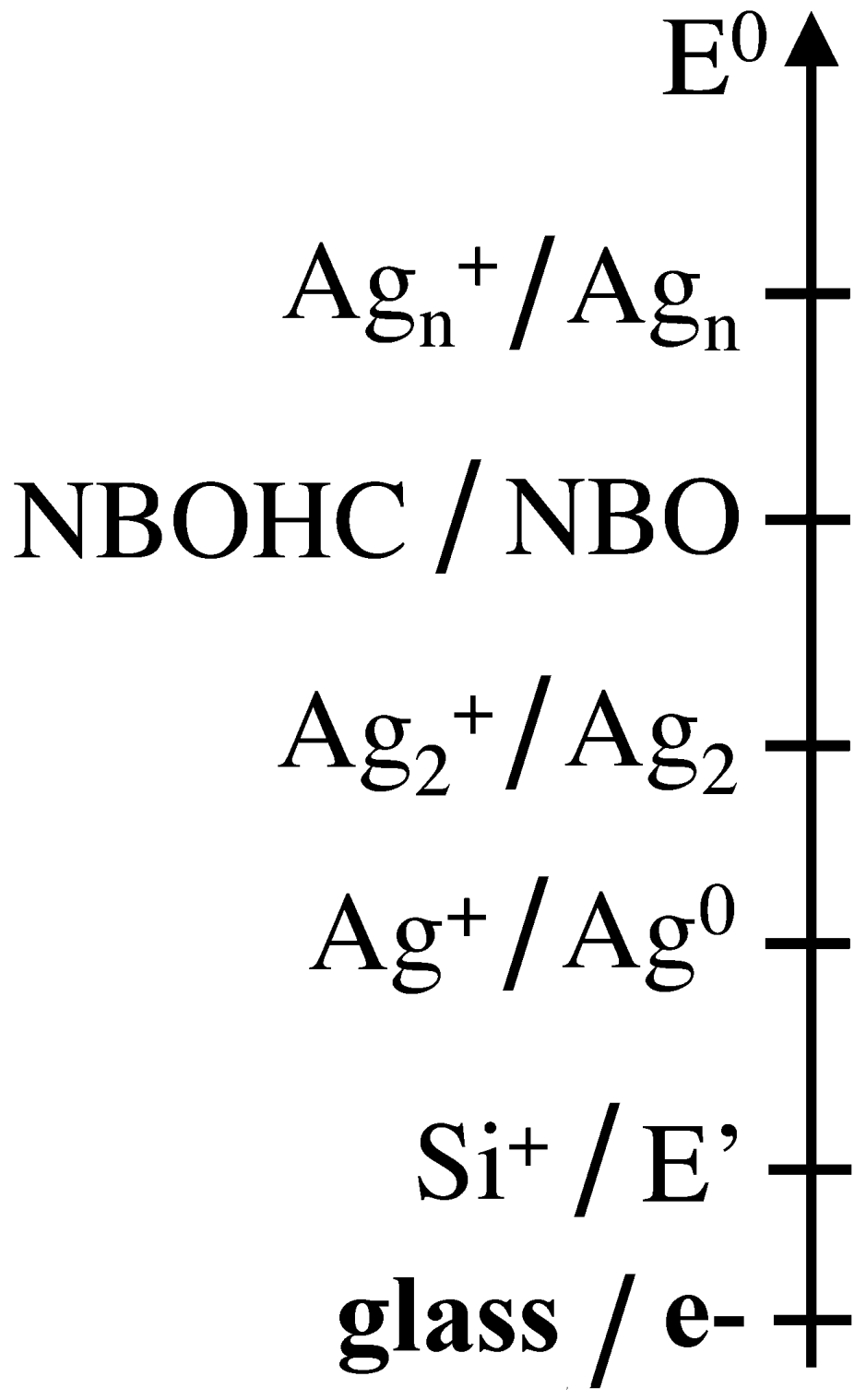

Fig. 17

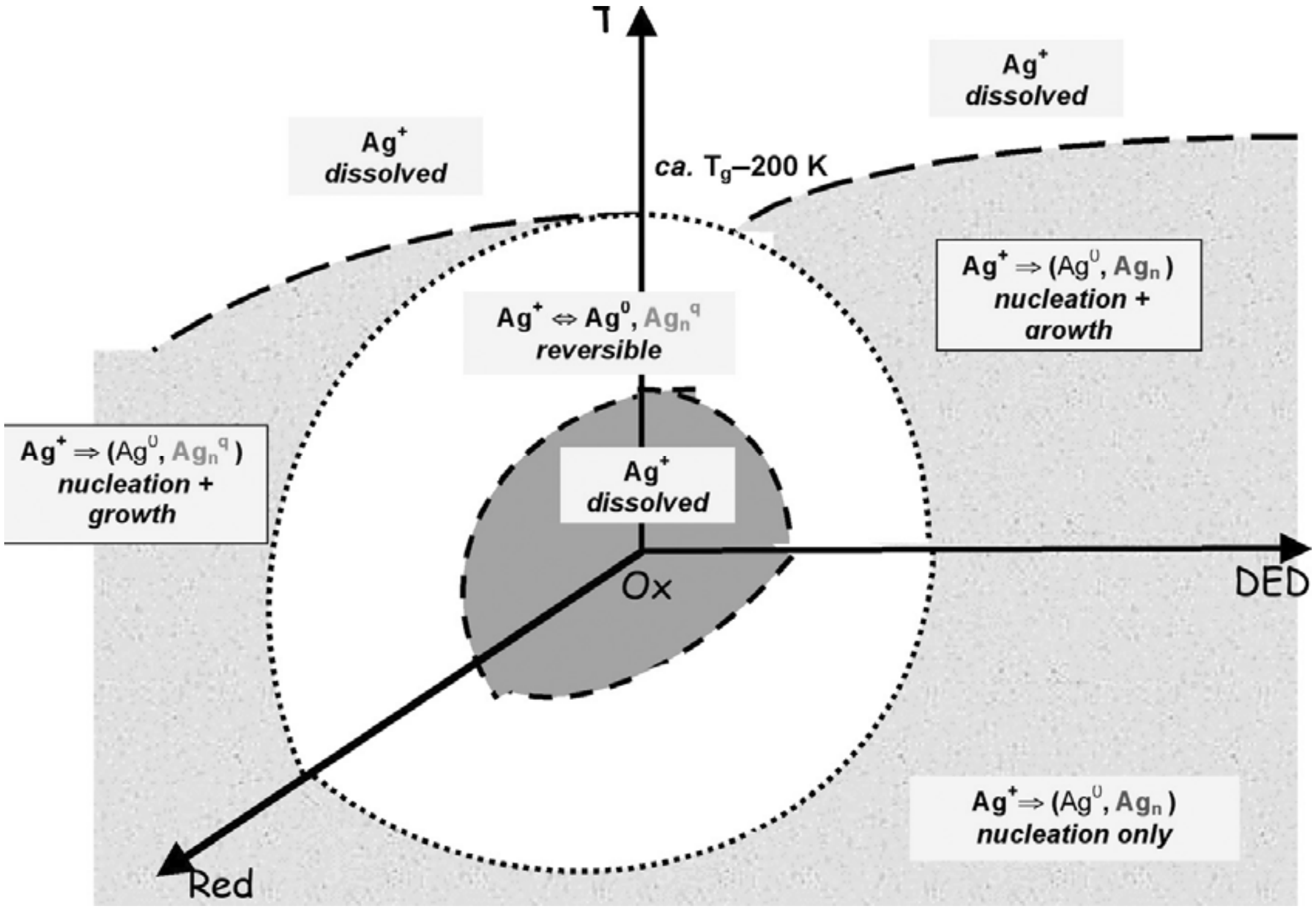

Fig. 18

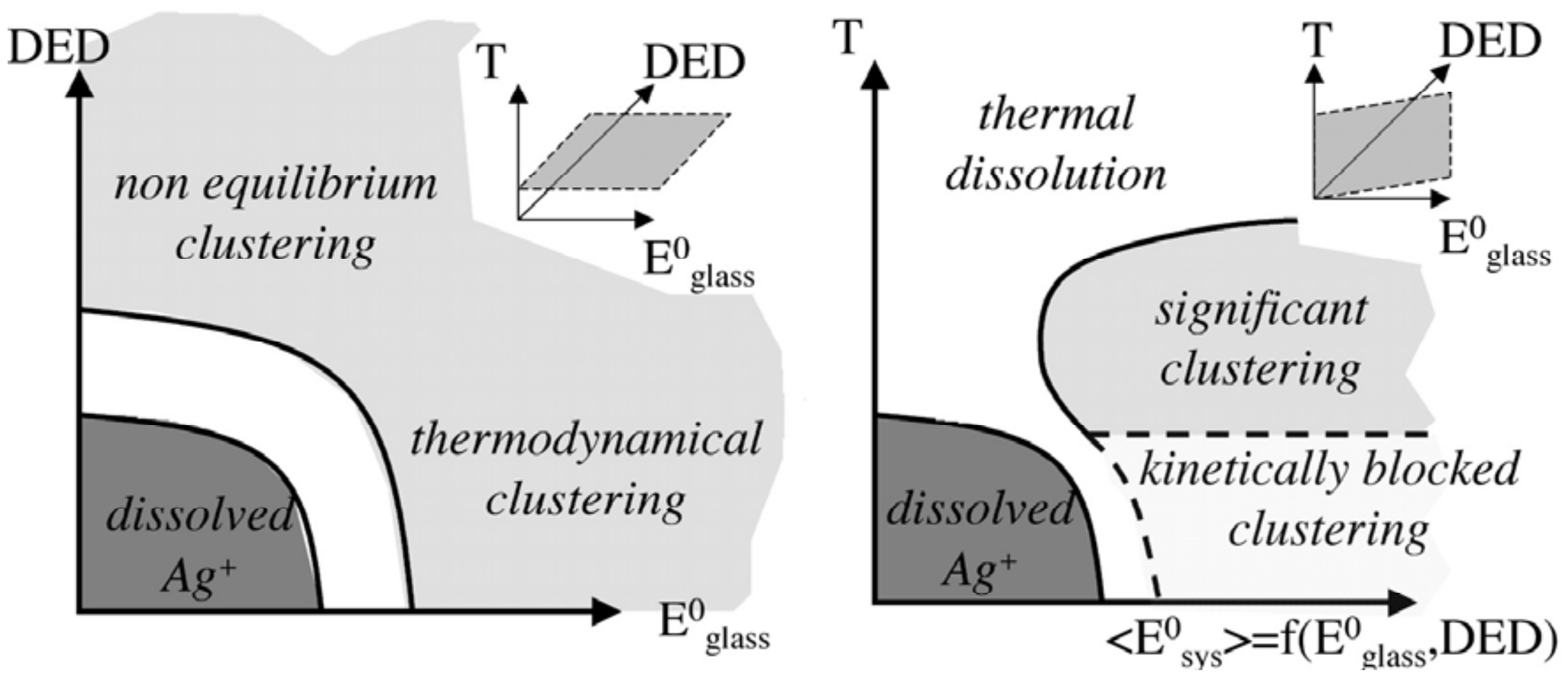